\documentclass{article}
\usepackage{mathtext}
\usepackage{amssymb}
\usepackage{graphics,graphicx}
\usepackage{color}
\usepackage{amsbsy}
\usepackage[left=2.5cm,right=2.5cm,top=2.5cm,bottom=2.5cm,bindingoffset=0cm]{geometry}
\usepackage{amsmath}
\usepackage{ccaption}
\usepackage{epstopdf}
\usepackage{ dsfont }
\DeclareMathOperator{\arccot}{arccot}
\usepackage{soul}
\usepackage{verbatim}
\usepackage{ dsfont }

\usepackage{cite}

\title{On admissible steady-state regimes of crack propagation in a square-cell lattice.}

\author{Nikolai Gorbushin, Gennady Mishuris
\\
{\it Department of Mathematics,
Aberystwyth University, }
\\ {\it Ceredigion SY23 3BZ, Wales, UK}
}
\date{}

\begin{document}
\maketitle

\begin{abstract}
In the present work the authors revisit a classical problem of crack propagation in a lattice. Authors investigate the questions concerning possible admissible steady-state crack propagations in an anisotropic lattice. It was found that for certain values of contrast in elastic and strength properties of a lattice the stationary crack propagation is impossible. Authors also address a question of possible crack propagation at low velocity.

\end{abstract}


\section{Introduction}

The difference in the radiation of mechanical waves in discrete and continuum solids effects on crack propagation in them. It seems that the first one who noticed that was Thomson in his work \cite{thomson1971}. Thomson used the simple model of two horizontal chains of oscillators linked together in vertical direction. It was shown that for the unstable crack propagation in one-dimensional discrete structure one should apply the load almost as twice as the load needed to break one link following Griffith's fracture criterion.  It was he who proposed the term "lattice trapping" characterising such a phenomenon.

Somehow later and independently from Thomson, Slepyan came out with the solution for the problem of a steady-state fault propagation in one-dimensional \cite{slepyan1984} and two-dimensional \cite{kulakhmetova1984} discrete structures. It was presented that while the fracture happens in the discrete structure not only the fracture energy is released but also the energy carried by the waves that have a crack tip as their source. Such fracture waves are spotted in the experiments \cite{rosakis1999} and numerical simulations \cite{gao2001}. Slepyan's approach was successfully applied for the study of fracture problems of various structures such as bridged crack propagation \cite{mishuris2008}, chain with non-local interaction \cite{gorbushin2016}, lattice with a structured interface \cite{mishuris2009}, lattice fracture with dissipation \cite{slepyan2012}. The method was also used to analyse the problems of phase transformations \cite{slepyan2005,truskinovsky2005}.

The most of interest in the study of crack propagation is focused on the analysis of their stability. Experimental data reveal the correlation between the crack speed and the patterns on the fracture surface \cite{ravi1998}. Based on the triangular lattice models and the above mentioned approach there were attempts to define the ranges of the crack speeds that correspond to instabilities \cite{fineberg1999,pechenik2002}. This, however, was also done by the consideration of the continuum solids \cite{willis1995,movchan1995,piccolroaz2010}.

It should be pointed out that the computation of the dynamical crack propagation problems are time-consuming for both continuum and discrete solids. Particularly, the steady-state regime that is detected experimentally \cite{fineberg1999} can be compared with the analytical solution based on the mentioned method. The application of the analytical technique allows to obtain loading parameters that lead to a steady-state without heavy computations. Moreover, the numerical simulation of the continuum models require the discretisation of spatial and temporal domains. Due to this one may wonder whether the achieved effects are valid for the true continuous model or influenced by the discretisation. For example, FEM analysis with implemented cohesive zone elements shoes the oscillations of stress fields close to a crack tip in time \cite{xu1994}.

One of the major theoretical results obtained by Slepyan is the dependence of the energy release rate on the crack speed. Slepyan \cite{slepyan2012} and Marder \cite{fineberg1999} marked the crack speeds that possibly correspond to the stable steady-state crack propagation.

In the last decades there was a significant progress in the study of dynamical crack growth in discrete media. For instance, different types of loads were considered and the peculiarities caused by them were studied. Among the list of the examples we can name the energy flux from infinity \cite{slepyan1984,kulakhmetova1984,mishuris2008,gorbushin2016}, waves of a certain frequencies coming from infinity \cite{mishuris2009}, constant displacement of lattice edges (in case of finite lattice width) \cite{fineberg1999,pechenik2002}. Notice that the problem itself is non-linear and superposition principle does not work. Furthermore, numerical simulations showed that there exist regimes that are not compatible with steady-state ones \cite{fineberg1999,pechenik2002} and the linear analysis of stability was performed \cite{marder1993}. Apart from that there were numerically observed \cite{mishuris2009} and theoretically explained later on in \cite{slepyan2010} so-called clustering quasi steady-state regime, where in average the crack moves regularly but within the period (cluster) it is not seen. Another periodic process,forerunning, was numerically studied \cite{slepyan2015}. It characterises by the periodic appearing of the damaged region ahead of the main running crack. Such effect is presented in experiments when the crack movement is supported by the developing microcracks ahead of it \cite{ravi1984}. Nevertheless, there are still many details remained untouched. So, after the prediction of possible stable steady-state regimes Marder wrote \cite{marder1995}: "The complete story has yet to be worked out".

In the present work we try to answer on the some of the addressed questions. Namely, is it possible to observe stable steady-state crack propagation at low speeds which was shown in \cite{gorbushin2016,gorbushin2017,mishuris2014}. In these papers it was shown that the anisotropy in the properties has an impact on the obtaining such an effect which was numerically confirmed. It should be stressed that Slepyan's approach deals with discrete difference problem of moving defect but not the original problem. Thus, the obtained steady-state regimes should be verified and their specifics should be analysed (see clustering above). Works \cite{gorbushin2017,mishuris2014} referred to the one-dimensional cases but the lattice problems and these two problems have meaningful differences.

We consider a crack propagation in a square-cell lattice with different elastic properties in its orthogonal directions. We derive the solution of the problem and analyse the steady-state regimes. The solution is gained similarly to \cite{slepyan2012} but with the additional information useful for the current study. We demonstrate the peculiarities brought by anisotropy and point out details that were not mentioned previously in the literature (admissible regimes, the displacement in the direction perpendicular to the crack).

\section{Analytical solution of the problem.}
\subsection{Mathematical formulation of the problem}
\begin{figure}[h!]
\center{\includegraphics[scale=0.8]{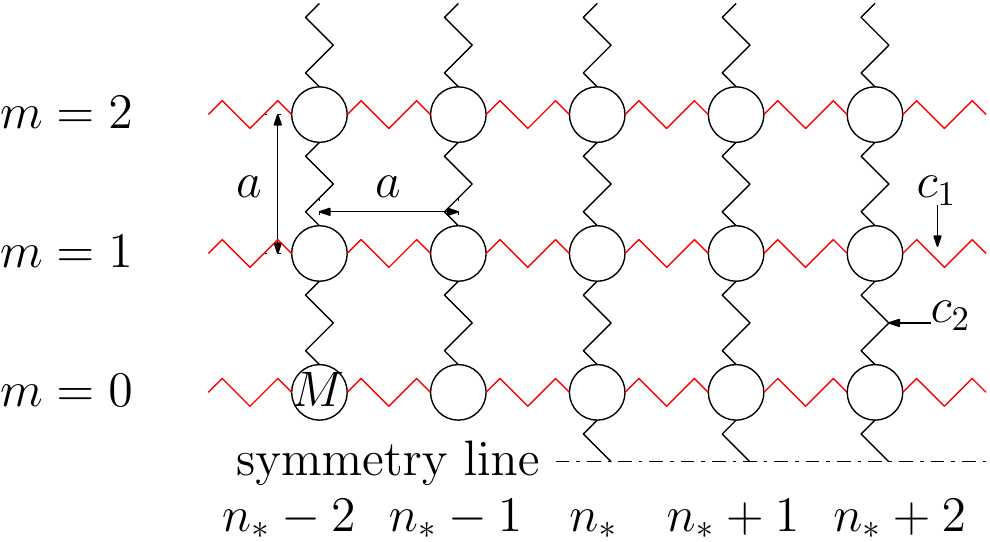}}
\captiondelim{. }
\caption[ ]{Infinite chain of oscillators with equal masses $M$ connected together by linear springs of stiffness $c_1$ in a horizontal direction and $c_2$ in a vertical direction. The crack position is defined by oscillator with index $n_*$. The vertical springs of the stiffness $c_2$ along the symmetry line break while the crack moves. $a$ is an equilibrium distance between the oscillators.}
\label{fig:Infinite Lattice}
\end{figure}

We proceed to the consideration of crack propagation in 2-d lattice shown in fig.\ref{fig:Infinite Lattice}.

The derivation of the solution can be found \cite{slepyan2012} and here we present the key points of it with the extensions for the further analysis. The governing equations for this problem are:
\begin{equation}
m=0:
\begin{array}{l}
M\ddot{u}_{0,n}=c_1(u_{0,n+1}+u_{0,n-1}-2u_{1,n})-2c_2u_{0,n}+c_2(u_{1,n}-u_{0,n}), \,n\geq n_*,\\[3mm]
M\ddot{u}_{0,n}=c_1(u_{0,n+1}+u_{0,n-1}-2u_{0,n})+c_2(u_{1,n}-u_{0,n}),\quad n< n_*,
\end{array}\\
\label{eq:LatticeProblemInitial}
\end{equation}
\begin{equation}
m\geq1:\quad\quad
M\ddot{u}_{m,n}=c_1(u_{m,n+1}+u_{m,n+1}-2u_{1,n})+c_2(u_{m+1,n}+u_{m-1,n}-2u_{m,n}),
\label{eq:LatticeProblemInitial_1}
\end{equation}
where $u_{m,n}=u_{m,n}(t)$ is an out of plane displacement of an oscillator in $m$-th horizontal layer and $n$-th vertical layer of a lattice.

The fracture condition for this problem is stated as:
\begin{equation}
u_{0,n_*}=u_c,
\label{eq:FractureCondition}
\end{equation}
\begin{equation}
u_{m,n}<u_c,\quad n>n_*.
\label{eq:FractureCondition_1}
\end{equation}

\subsection{Solution for steady-state crack propagation in lattice}
We switch to moving coordinate system with the origin at the crack tip by the mean of change of variable:
\begin{equation}
\eta=n-vt,\quad v=const
\label{eq:eta}
\end{equation}

In \eqref{eq:eta} we take into account that the crack moves steadily with a constant speed $v$ and we assume that the coordinate $\eta$ can be treated as continuous. The crack speed is assumed to be limited by the value $v_c$:
\begin{equation}
v<v_c=\sqrt{\frac{c_1}{M}}.
\label{eq:CriticalSpeed}
\end{equation}

The quantity $v_c$ is equal to Rayleigh speed in the broken part of the chain. We also assume that the solution is a function of $\eta$ only in such a crack propagation regime:
We assume that the solution can be presented as:
\begin{equation}
u_{m,n}(t)=u_m(\eta)
\end{equation}
Fracture conditions \eqref{eq:FractureCondition} and \eqref{eq:FractureCondition_1} are modified as:
\begin{equation}
u_{0}(0)=u_c,
\label{eq:FractureCondition_eta}
\end{equation}
\begin{equation}
u_{m}(\eta)<u_c,\quad \eta>0.
\label{eq:FractureCondition_eta_1}
\end{equation}
Equations of motion \eqref{eq:LatticeProblemInitial} become:
\begin{equation}
\begin{gathered}
Mv^2\frac{d^2}{d\eta^2}u_0(\eta)=c_1(u_0(\eta+1)+u_0(\eta-1)-2u_0(\eta))-2c_1u_0(\eta)H(\eta)
+c_2(u_1(\eta)-u_0(\eta)),\,m=0,\\
Mv^2\frac{d^2}{d\eta^2}u_m(\eta)=c_1(u_m(\eta+1)+u_m(\eta-1)-2u_m(\eta))+c_2(u_{m+1}(\eta)
+u_{m-1}(\eta)-2u_m(\eta)),\,m\geq1.
\end{gathered}
\label{eq:LatticeProblemEta}
\end{equation}

We would like to perform the Bloch-Floquet analysis of the problem and introduce the notations for the dispersion relations that appear to be useful for the present study.

\subsection{Dispersion relations}

\begin{figure}[h!]
\minipage{0.45\textwidth}
\center{\includegraphics[width=\linewidth] {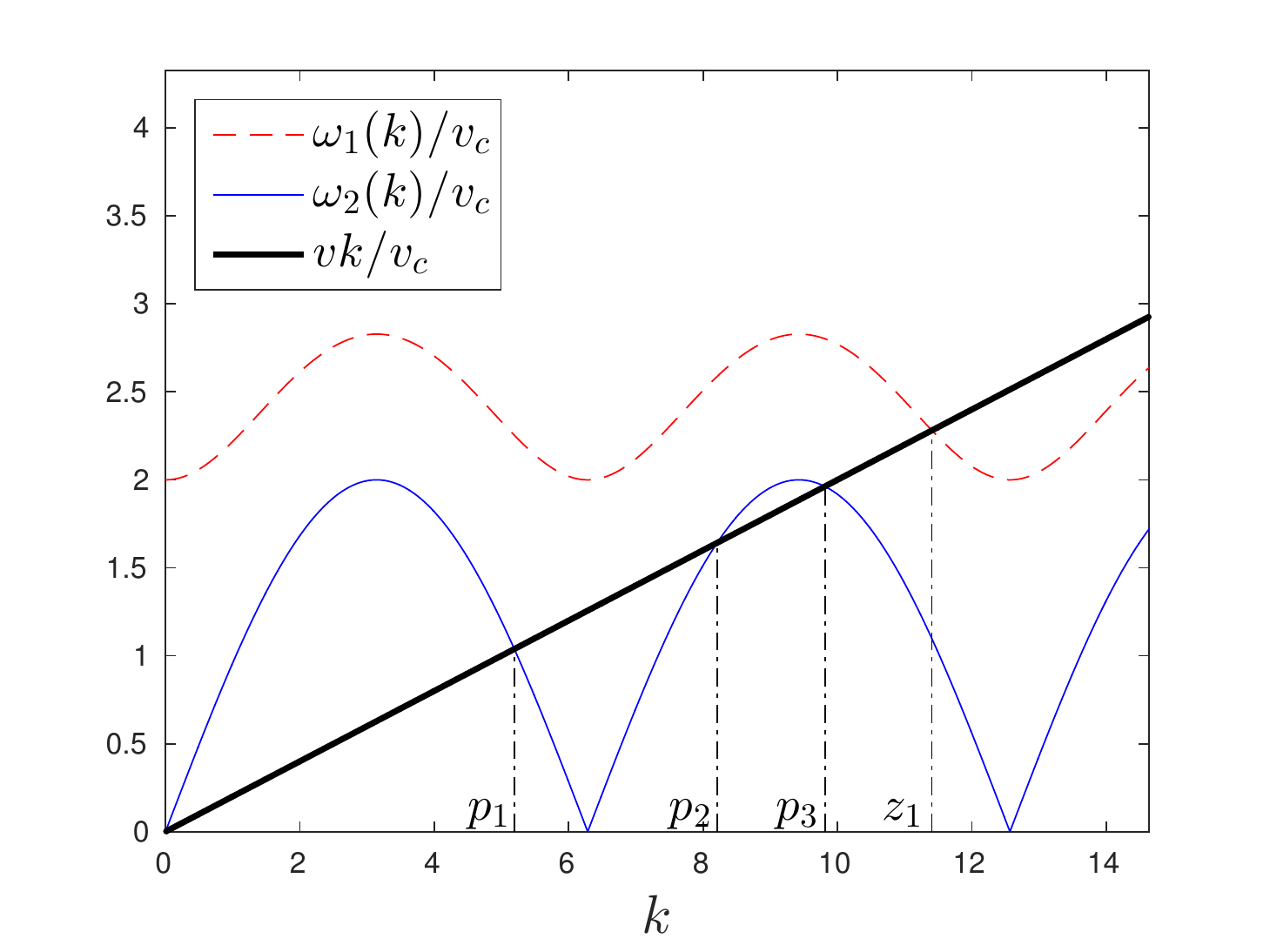} \\ a)}
\endminipage
\hfill
\minipage{0.45\textwidth}
\center{\includegraphics[width=\linewidth] {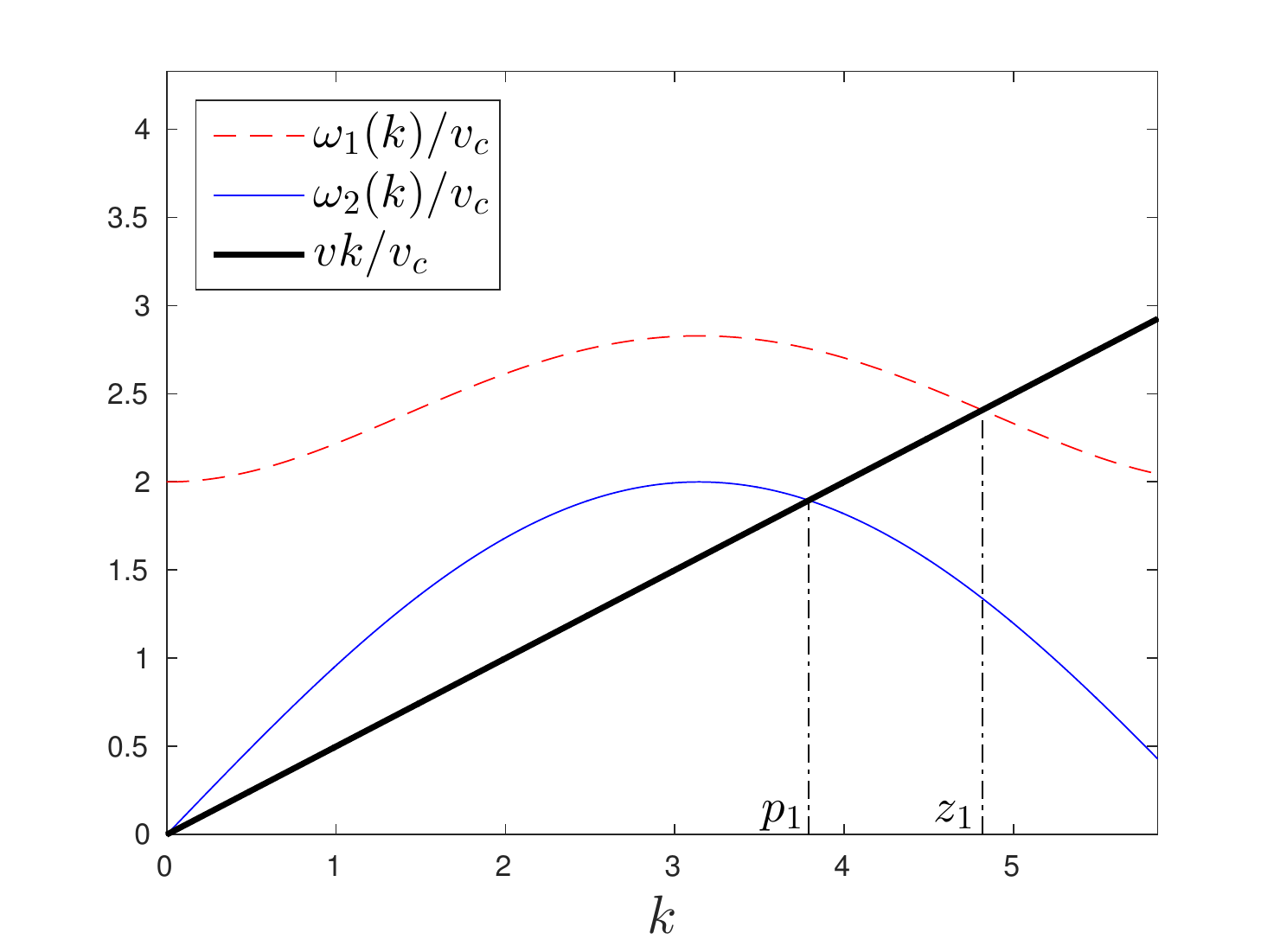} \\ b)}
\endminipage
\hfill
\captiondelim{. }
\caption[ ]{Dispersion relations for different values of crack speed and $c_2=c_1$: a) $v=0.2v_c$, b) $v=0.5v_c$.}
\label{fig:DispersionRelations}
\end{figure}

It is convenient to introduce notations for the dispersion relations that appear in the problem. The first relation is:
\begin{equation}
\omega_1^2(k)=4v_c^2\sin^2{\left(\frac{k}{2}\right)}+4\omega_0^2,\quad \omega_0^2=\frac{c_2}{M},
\label{eq:DispersionRelationLattice_Intact}
\end{equation}
whereas the second relation is:
\begin{equation}
\omega_2^2(k)=4v_c^2\sin^2{\left(\frac{k}{2}\right)}
\label{eq:DispersionRelationLattice_Broken}
\end{equation}
Relations \eqref{eq:DispersionRelationLattice_Intact}, \eqref{eq:DispersionRelationLattice_Broken} define the characteristics of waves in horizontal direction. The plots of dispersion relations are shown in fig. \ref{fig:DispersionRelations} for two values of crack speed $v=0.2v_c$ and $v=0.5v_c$.

In the fig.\ref{fig:DispersionRelations} constants $z_j$ and $p_j$ are positive roots of the following equations:
\begin{equation}
\begin{gathered}
\omega_1(q_j)-vz_j=0,\quad j=1...Z,\\
\omega_2(p_j)-vp_j=0,\quad j=1...P,
\end{gathered}
\end{equation}
where $Z$ and $P$ are their total number, respectively.
\subsection{Description of problem in vertical direction}
The last equations in \eqref{eq:LatticeProblemEta} is:
\begin{equation}
Mv^2\frac{d^2}{d\eta^2}u_m(\eta)=c_1(u_m(\eta+1)+
u_m(\eta-1)-2u_m(\eta))+c_2(u_{m}(\eta+1)+u_{m-1}(\eta)-2u_{m}(\eta)),\quad m\geq1.
\end{equation}
The application of Fourier transform to this equation gives:
\begin{equation}
\left[(0+ikv)^2+\omega_2^2(k)+2\omega_0^2\right]U_m(k)=\omega_0^2(U_{m+1}(k)+U_{m-1}(k)).
\label{eq:FourierInsideLattice}
\end{equation}
From now on we mean:
\begin{equation}
(0\pm ik)=\lim_{s\to0+}(s\pm ik)
\label{eq:Limit_s}
\end{equation}

Taking advantage from the fact that the coefficients in the system of linear equations do not depend on the value of $n$ (or what is equivalent that the material properties of the system do not change in the direction perpendicular to the crack line we assume existence of a function $\lambda(k)$ such that:
\begin{equation}
U_m(k)=\lambda^{m}(k)U_0(k),\quad m\geq1,
\label{eq:LambdaIntroduction}
\end{equation}
where the function $\lambda(k)$ should satisfy the following condition
\begin{equation}
|\lambda(k)|\le1.
\label{eq:Lambda_condition}
\end{equation}
If the measure of the set of the $k$ where the equality is satisfied is equal to zero:
\begin{equation}
\mbox{mes}\, \mathcal{K}=\{k\in \mathbb{R}: |\lambda(k)|=1\},
\label{eq:Lambda_measure}
\end{equation}
then the solution of the original problem $u_{m,n}$ will vanish at infinity ($m\to\infty$). Otherwise the waves profile remains visible for any $m$.

Substituting \eqref{eq:LambdaIntroduction} into \eqref{eq:FourierInsideLattice} after some algebra one
obtains a quadratic equation to determine the function $\lambda(k)$:
\begin{equation*}
\lambda^2(k)-\frac{1}{\omega_0^2}\left((0+ikv)^2+\omega_2^2(k)+2\omega_0^2\right)\lambda(k)+1=0,
\end{equation*}
which naturally has two solutions $\lambda_j(k)$, $k=1,2$:
\begin{equation}
\lambda_1(k)=\lambda(k),\quad \lambda_1(k)=\frac{1}{\lambda(k)},
\label{eq:Lambda}
\end{equation}
where we have introduced a new function:
\begin{equation}
\lambda(k)=\frac{\sqrt{(0+ikv)^2+\omega_1^2(k)}-\sqrt{(0+ikv)^2+\omega_2^2(k)}
}{\sqrt{(0+ikv)^2+\omega_1^2(k)}+\sqrt{(0+ikv)^2+\omega_2^2(k)}
}
\label{eq:Lambda_final}
\end{equation}

Note that the square roots in the formula are chosen in such a way that for any $s>0$ from the limit in \eqref{eq:Limit_s}
they represent the same branches and thus are continuous. Following the same reasoning as in \cite{slepyan2012} we conclude that the root $|\lambda(k)|\leq 1,\, k\in \mathbb{R}$ and, consequently, only the first root $\lambda_1(k)$ should be taken.

By straightforward substitution it is easy to check:
\begin{equation}
\lambda(p_j)=1, \quad j=1,2,...,P,
\label{eq:Lambda_3}
\end{equation}
and
\begin{equation}
\lambda(z_j)=-1, \quad j=1,2,...,Z.
\label{eq:Lambda_4}
\end{equation}
Moreover, $\lambda(0)=1$. Note that there are two countable sets $\mathcal{K}_j=\{k\in \mathbb{R}: \lambda(k)=(-1)^{j}\}$
where the condition \eqref{eq:Lambda_measure} satisfies. The other cases when $|\lambda(k)|=1$ but $\lambda\in  \hspace{-4mm}\diagup \, \mathbb{R}$ and thus
\begin{equation}
\lambda(k)=e^{i\phi_\lambda(k)}, \quad \phi_\lambda(k)\in \mathbb{R},
\label{eq:Lambda_arg}
\end{equation}
appear only when one of the square roots are pure imaginary while the other is real and
\begin{equation}
\tan\frac{\phi_\lambda(k)}{2}=-i\frac{\sqrt{(0+ikv)^2+\omega_1^2(k)}}{\sqrt{(0+ikv)^2+\omega_2^2(k)}},\quad
\left(\omega_1^2(k)-k^2v^2\right)\left(\omega_2^2(k)-k^2v^2\right)<0.
\label{eq:Lambda_arg_fun}
\end{equation}

Function $\phi_\lambda(k)$ should be chosen in such a manner that $\mbox{arg}\lambda(k)$ is a continuous along the entire the real axis. One can prove that the following properties of the constructed function $\lambda(k)$:
\begin{equation}
|\lambda(k)|=|\lambda(-k)|,\quad  \text{arg}\lambda(k)=-\text{arg}\lambda(-k),\quad\text{for } k\in\mathbb{R}.
\label{Properties_lambda}
\end{equation}
We can also write the asymptotic behaviour of the function at infinity:
\begin{equation}
\lambda(k)\sim-\frac{2\omega_0^2}{v^2k^2},\quad k\to\infty.
\label{eq:Asymptotics_Lambda_Infty}
\end{equation}
The asymptotic behaviour of $\lambda(k)$ at zero comes from \eqref{eq:Asymptotics_Lattice_L_zero}:
\begin{equation}
\lambda(k)=1-\sqrt{\frac{v_c^2-v^2}{\omega_0^2}}\sqrt{0+ik}\sqrt{0-ik}+o(k),\quad k\to0.
\label{eq:Asymptotics_Lambda_Zero}
\end{equation}
Note also that $\lambda(k)$ takes only real value for $-p_1<k<p_1$.

\begin{figure}[h!]
\minipage{0.45\textwidth}
\center{\includegraphics[width=\linewidth] {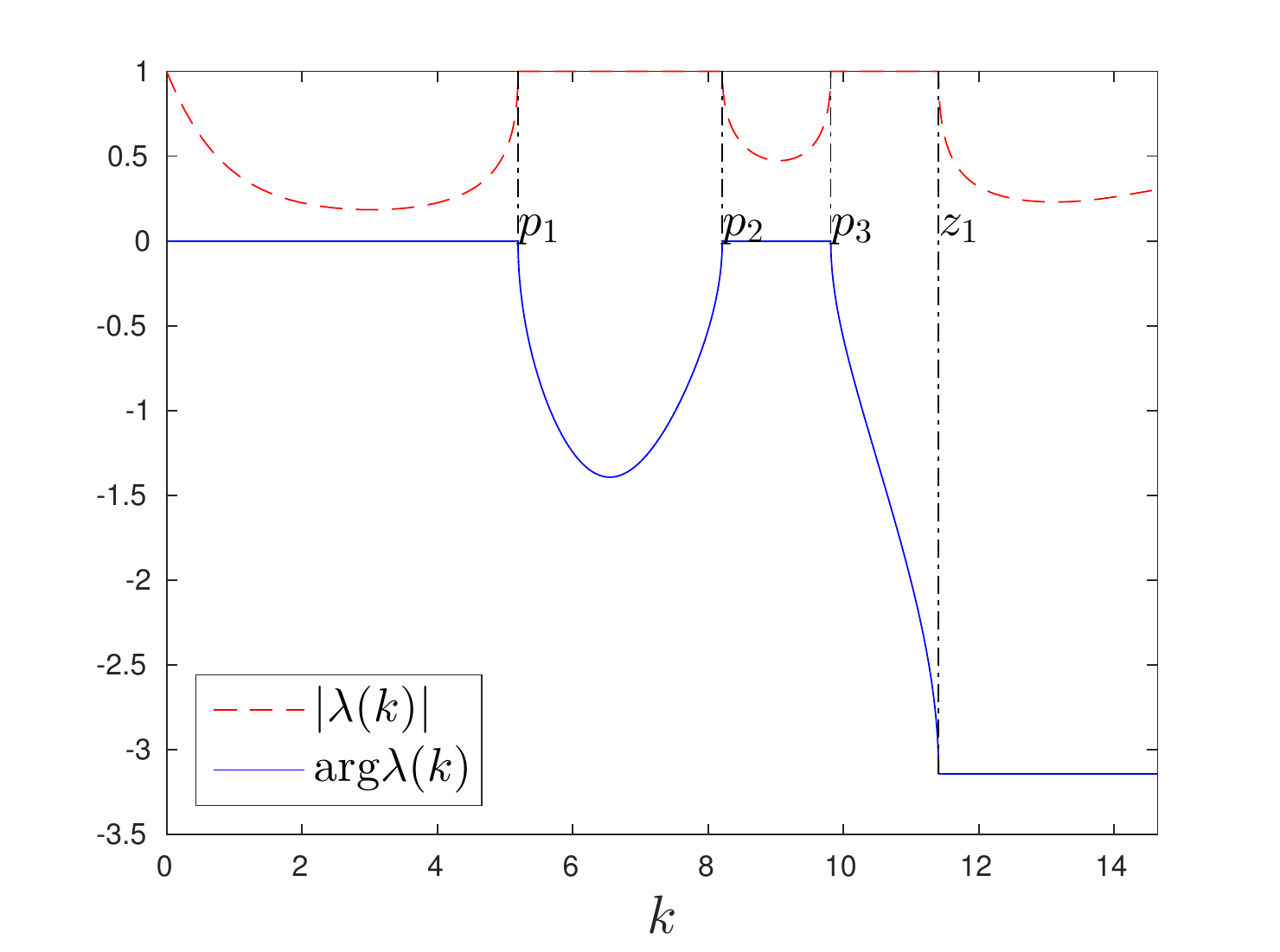} \\ a)}
\endminipage
\hfill
\minipage{0.45\textwidth}
\center{\includegraphics[width=\linewidth] {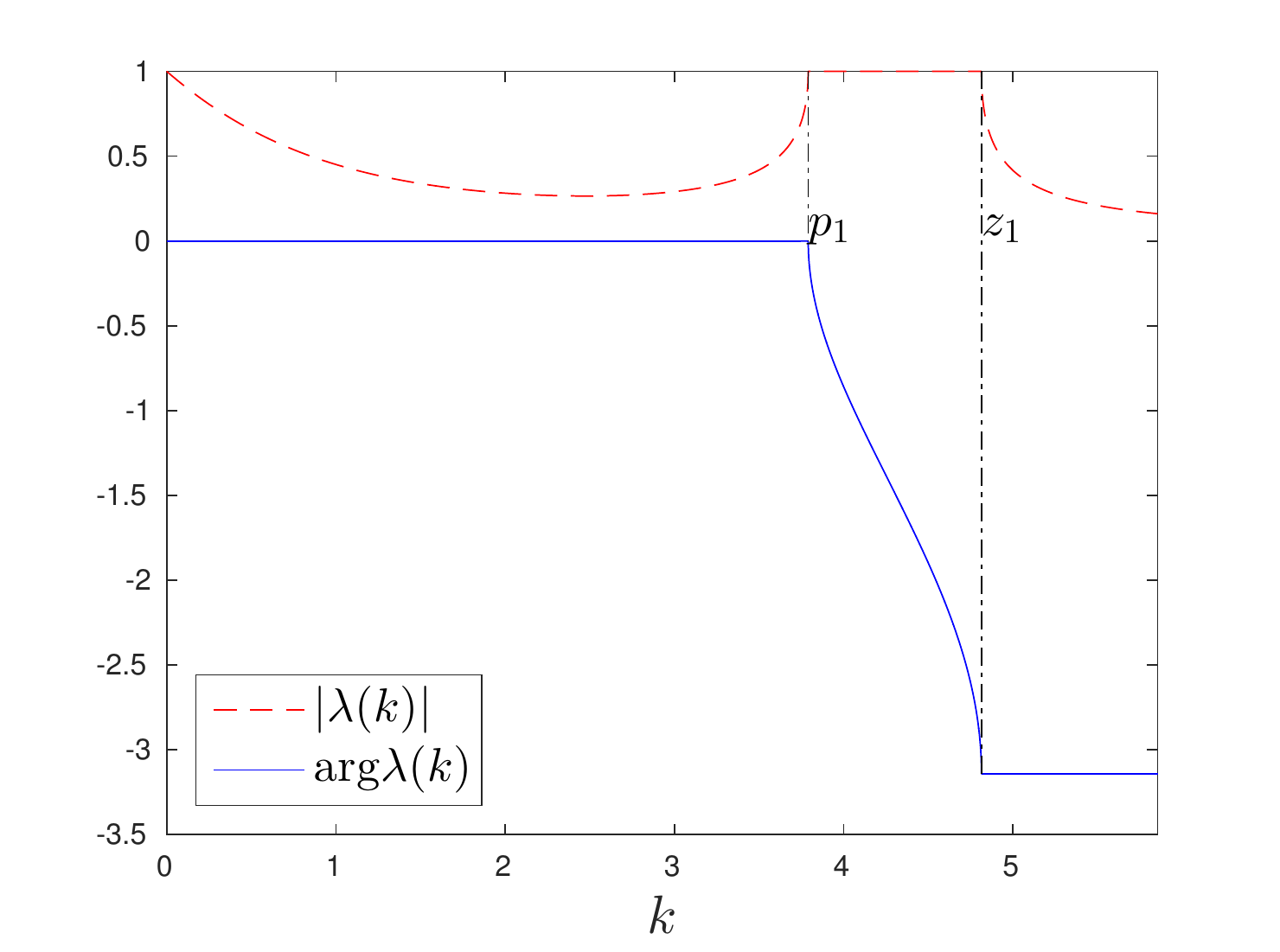} \\ b)}
\endminipage
\hfill
\captiondelim{. }
\caption[ ]{Absolute value and argument of $\lambda(k)$ for different values of crack speed and $c_2=c_1$: a) $v=0.2v_c$, b) $v=0.5v_c$.}
\label{fig:Labda_abs_arg}
\end{figure}

To illustrate the behaviour of function $\lambda(k)$, we present in fig.\ref{fig:Labda_abs_arg}  the graphs of
$|\lambda(k)|$ and $\mbox{arg}(\lambda(k))$ for the cases $v=0.2v_c$ and $v=0.5v_c$ for positive values of $k$. In this figures it can be explicitly seen that $|\lambda(k)|<1$ only when $(\omega_1^2(k)-(kv)^2)(\omega_1^2(k)-(kv)^2)<0$ which is supported by the dispersion relations in fig. \ref{fig:DispersionRelations}.

Note that the $\mbox{arg}\lambda(k)$ is an odd continuous function satisfying the properties for positiuve values of its argument ($k>0$)
\[
\lim_{k\to\infty}\mbox{arg}\lambda(k)=\pi,\quad \mbox{and}\quad 0<\mbox{arg}\lambda(k)<\pi \quad \mbox{only if}\quad |\lambda(k)|=1.
\]
Another conclusion immediately follows from this analysis is that the waves profile does not vanishes at infinity ($m\to\infty$).

\subsection{Derivation of Wiener-Hopf type equation}
Let us start from the equations \eqref{eq:LatticeProblemEta}. The use of Fourier transform reduces the problem to equations:
\begin{equation}
\begin{gathered}
\Big((0+ikv)^2+\omega_2^2(k)+2\omega_0^2\Big)U^+(k)
+\Big((0+ikv)^2+\omega_2^2(k)\Big)U^-(k)=\omega_0^2\Big( U_1(k)-U(k)\Big),\quad m=0\\
\Big((0+ikv)^2+\omega_2^2(k)+2\omega_0^2\Big)U_m(k)=\omega_0^2\Big(U_{m+1}(k)-U_{m-1}(k)\Big),\quad m\geq1.
\end{gathered}
\label{eq:LatticeFourierTransform}
\end{equation}
where the notations are:
\begin{equation}
\begin{gathered}
U(k)=U_0(k)=\int_{-\infty}^{\infty}u_0(\eta)e^{ik\eta}\,d\eta=U^+(k)+U^-(k),\quad U^{\pm}(k)=\int_{-\infty}^{\infty}u_0(\eta)H(\pm\eta)e^{ik\eta}\,d\eta\\
U_m(k)=\int_{-\infty}^{\infty}u_m(\eta)e^{ik\eta}\,d\eta,\quad m\geq1.
\end{gathered}
\end{equation}

With the introduced function $\lambda(k)$ in \eqref{eq:LambdaIntroduction} the equations for $m\geq1$ are already satisfied while the equation for $m=0$ is reduced to:
\begin{equation}
L(k)U^+(k)+U^-(k)=0
\label{eq:WienerHopf_Lattice_Initial}
\end{equation}
Kernel function $L(k)$ in this case is defined as:
\begin{equation}
L(k)=\sqrt{\frac{(0+ikv)^2+\omega_1^2(k)}{(0+ikv)^2+\omega_2^2(k)}},
\label{eq:FunctionL_Lattice}
\end{equation}
where the functions $\omega_{1,2}(k)$ are defined in \eqref{eq:DispersionRelationLattice_Intact},\eqref{eq:DispersionRelationLattice_Broken} and function $\lambda(k)$ is presented in \eqref{eq:Lambda_final}.
One can also notice that function $L(k)$ has the following properties:
\begin{equation}
|L(k)|=|L(-k)|,\quad \text{arg}L(k)=-\text{arg}L(-k),\quad\text{for } k\in\mathbb{R}
\label{PropertiesL}
\end{equation}
The asymptotic behaviour is given as:
\begin{equation}
L(k)=1-\frac{2\omega_0^2}{v^2k^2}+O(k^{-4}),\quad  k\to\infty,
\label{eq:Asymptotics_Lattice_L_infty}
\end{equation}
\begin{equation}
L(k)\sim
\frac{2\omega_0}{\sqrt{v_c^2-v^2}}\frac{1}{\sqrt{(0+ik)(0-ik)}}, \quad k\to0,
\label{eq:Asymptotics_Lattice_L_zero}
\end{equation}

\begin{figure}[h!]
\minipage{0.45\textwidth}
\center{\includegraphics[width=\linewidth] {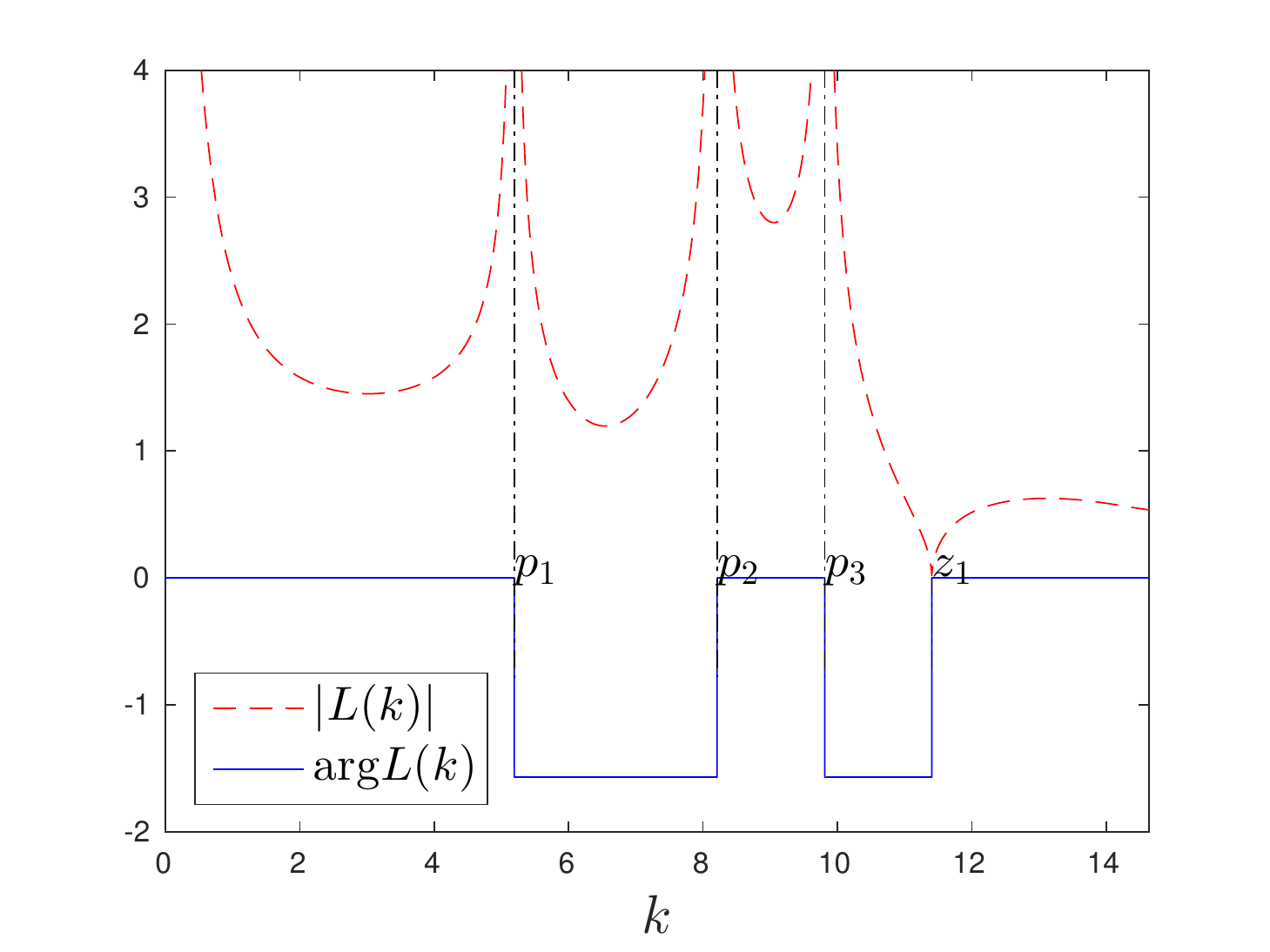} \\ a)}
\endminipage
\hfill
\minipage{0.45\textwidth}
\center{\includegraphics[width=\linewidth] {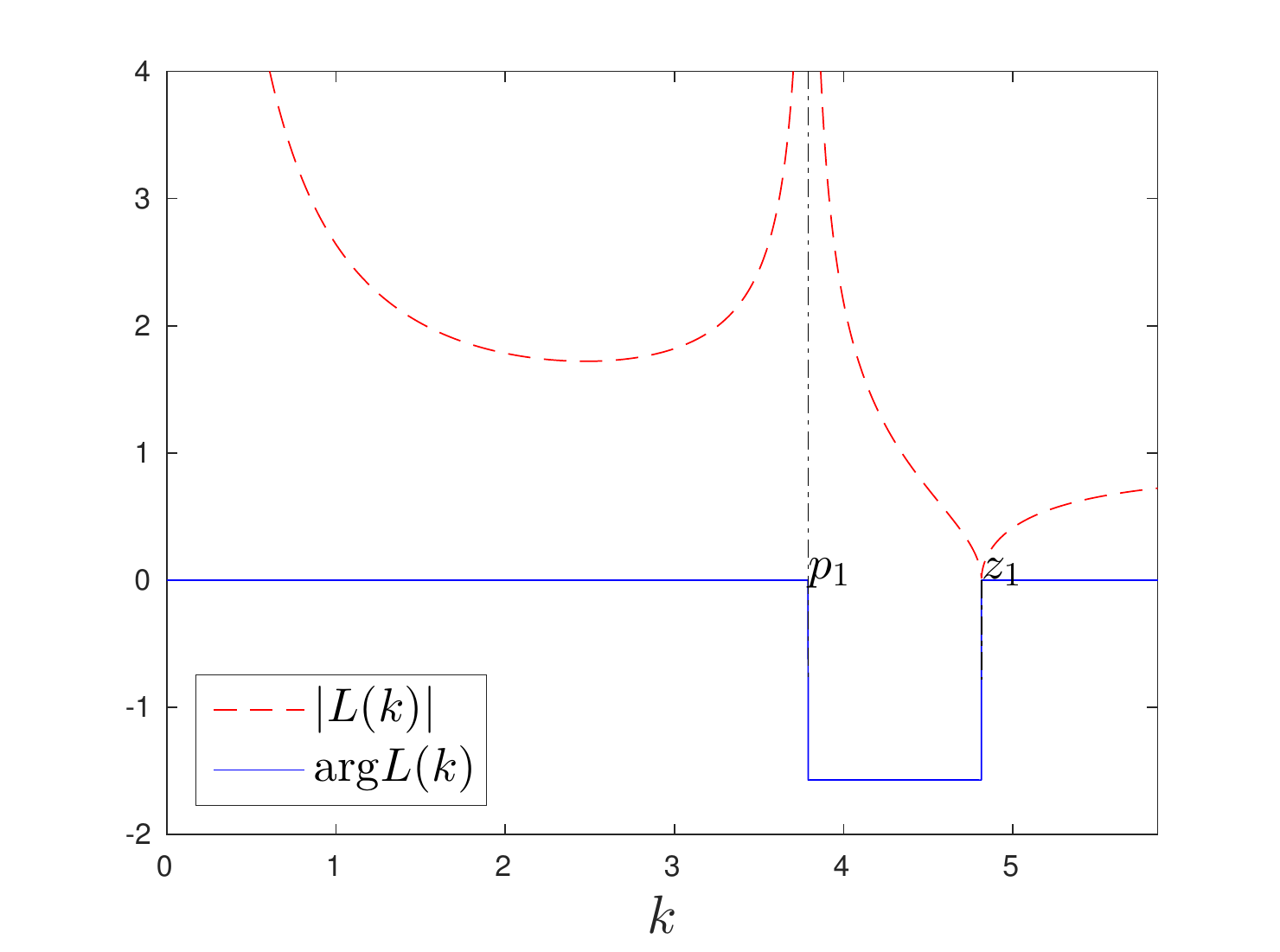} \\ b)}
\endminipage
\hfill
\captiondelim{. }
\caption[ ]{Absolute value and argument of $L(k)$ for different values of crack speed and $c_2=c_1$: a) $v=0.2v_c$, b) $v=0.5v_c$.}
\label{fig:L_abs_arg}
\end{figure}

Plots of absolute value and argument of $L(k)$ are presented in fig.\ref{fig:L_abs_arg} for $v=0.2v_c$ and $v=0.5v_c$. In these figures we observe that function $argL(k)$ is a stepwise. It experiences jumps of magnitude $\pi/2$ at poles and zeros of function $L(k)$.

\subsection{Solution of the Wiener-Hopf equation \eqref{eq:WienerHopf_Lattice_Initial}}

We can factorise the kernel function $L(k)=L^+(k)L^-(k)$ with use of the Cauchy type theorem together with Sokhotski-Plemelj relation as:
\begin{equation}
L^\pm(k)=\exp{\left(\pm\frac{1}{2\pi i}\int\limits_{-\infty}^{\infty}\frac{\text{Ln}{L(\xi)}}{\xi-k}\,d\xi\right)}.
\label{eq:Factorisation}
\end{equation}
The factors  $L^{\pm}(k)$ are analytic in the half-planes $\pm\text{Im}k>0$ and satisfy
the following asymptotic relations:
\begin{equation}
\begin{gathered}
L^{\pm}(k)-1\sim \pm i\frac{Q}{k},\quad k\to\infty,\\
Q=\frac{1}{\pi}\int\limits_{0}^{\infty}\text{log}{|L(\xi)|}\,d\xi.
\end{gathered}
\label{eq:Asymptotics_Lattice_L+-_infty}
\end{equation}
One can also estimate asymptotic behaviour of the factors near zero point with the use of \eqref{eq:Asymptotics_Lattice_L_zero} and \eqref{eq:Factorisation}:
\begin{equation}
\begin{gathered}
L^{\pm}(k)\sim R^{\pm1}\left(\frac{4\omega_0^2}{v_c^2-v^2}\right)^{1/4}\frac{1}{\sqrt{0\mp ik}}\left(1+(0\mp ik)S\right),\quad k\to0,\\[3mm]
R=\exp{\left(\frac{1}{\pi}\int\limits_0^\infty\frac{\text{Arg}L(\xi)}{\xi}\,d\xi\right)},\quad S=\frac{1}{\pi}\int\limits_0^\infty\frac{\log{|L(\xi)|}}{\xi^2}\,d\xi.
\end{gathered}
\label{eq:Asymptototics_Lattice_L+-_zero}
\end{equation}
Taking into an account the asymptotic relations the right part of the last equation \eqref{eq:WienerHopf_Lattice_Initial} can be modified \cite{slepyan2012}:
\begin{equation}
L^+(k)U^+(k)+\frac{1}{L^-(k)}U^-(k)=\frac{C}{0-ik}+\frac{C}{0+ik},
\label{eq:WienerHopf_Lattice_Final}
\end{equation}
where $C=const$. The solution for the Wiener-Hopf problem is:
\begin{equation}
U^+(k)=\frac{C}{0-ik}\frac{1}{L^+(k)},\quad U^-(k)=\frac{C}{0+ik}L^-(k)
\label{eq:SolutioFourierLattice}
\end{equation}
The displacement $u(\eta)$ is obtained in terms of inverse Fourier transform:
\begin{equation}
u_0(\eta)=\frac{1}{2\pi}\int_{-\infty}^{\infty}U^{\pm}(k)e^{-ik\eta}\,dk,\quad \pm\eta>0
\label{eq:SolutionLatticeInverseFourier}
\end{equation}
From \eqref{eq:Asymptotics_Lattice_L+-_infty} and \eqref{eq:Asymptototics_Lattice_L+-_zero}  it follows that:
\begin{equation}
\begin{gathered}
U^{\pm}(k)\sim C\left(\pm\frac{i}{k}+\frac{Q}{k^2}\right),\quad k\to\infty,\\
U^+(k)=\frac{C}{R}\left(\frac{v_c^2-v^2}{4\omega_0^2}\right)^{1/4}\frac{1}{\sqrt{0-ik}}+o(1),\quad k\to+0,\\
U^-(k)=\frac{C}{R}\left(\frac{4\omega_0^2}{v_c^2-v^2}\right)^{1/4}\left(\frac{1}{(0+ik)^{3/2}}+\frac{S}{\sqrt{0+ik}}\right)+o(1),\quad k\to-0.
\end{gathered}
\label{eq:Asymptotics_Lattice_Fourier}
\end{equation}
First relations in \eqref{eq:Asymptotics_Lattice_Fourier} give the asymptotic behaviour of solution $u(\eta)$ at zero:
\begin{equation}
u_0(\eta)=C(1-Q\eta)+O(\eta^2),\quad \eta\to0,
\end{equation}
whereas from the last 2 expressions in \eqref{eq:Asymptotics_Lattice_Fourier} by means of contour integration we conclude:
\begin{equation}
\begin{gathered}
u_0(\eta)=\frac{C}{R}\left(\frac{v_c^2-v^2}{4\omega_0^2}\right)^{1/4}\frac{1}{\sqrt{\pi\eta}},\quad \eta\to\infty,\\
u_0(\eta)=\frac{C}{R}\left(\frac{4\omega_0^2}{v_c^2-v^2}\right)^{1/4}\left(2\sqrt{\frac{-\eta}{\pi}}+\frac{S}{\sqrt{-\pi\eta}}\right),\quad \eta\to-\infty
\end{gathered}
\end{equation}
Constant $C$ is found from the fracture criterion \eqref{eq:FractureCondition_eta}:
\begin{equation}
C=u_c
\end{equation}
With the help of \eqref{eq:LambdaIntroduction} we can find that:
\begin{equation}
u_m(\eta)=\frac{1}{2\pi}\int_{-\infty}^{\infty}\lambda^{m}(k)U(k)e^{-ik\eta}\,dk=\frac{1}{2\pi}\int_{-\infty}^{\infty}\lambda^{m}(k)[U^+(k)+U^-(k)]e^{-ik\eta}\,dk,\quad m\geq1.
\label{eq:SolutionLatticeInverseFourier_Layers}
\end{equation}

The asymptotic behaviour of function $u_m(\eta)$ when $\rho=\sqrt{m^2+\eta^2}=m\sqrt{1+\alpha^2}\to\infty$, assuming that $\eta=\alpha m$, is derived in Appendix. For convenience the polar coordinates were introduced:
\begin{equation}
m=\rho\sin{\theta},\quad \eta=\rho\cos{\theta},\quad
\cot{\theta}=\alpha,
\end{equation}
such that $\rho$ is a radial distance from the crack tip and $\theta$ is an angle between the radius vector and axis $\eta$, i.e. $0\leq\theta<\pi$ for the considered configuration shown in fig.\ref{fig:Infinite Lattice}. The result for is summarised in formula \eqref{eq:AngleFunction_2} which is given as:

\begin{equation}
u_m(\eta)\sim \frac{2u_c}{R}\left(\frac{4\omega_0^2}{v_c^2-v^2}\right)^{1/4}\!\!\sqrt{\frac{\rho}{2\pi}}\,\,\,\Phi(\theta,v),\quad \rho\to\infty,
\label{eq:Asymptotics_m}
\end{equation}
\begin{equation}
\Phi(\theta,v)=
\left(\sqrt{\cos^2\theta+\frac{v_c^2-v^2}{\omega_0^2}\sin^2\theta}- \cos\theta \right)^{1/2},\quad 0\le \theta<\pi,
\label{eq:AngleFunction}
\end{equation}
\section{Solution analysis}

In the works of Slepyan \cite{slepyan1984,slepyan2012} the energetic relations were derived. It was shown that the ratio between the local energy release rate and global energy release rate is:
\begin{equation}
\frac{G_0}{G}=R^2, \quad G_0=\frac{2c_1u_c^2}{a}
\label{eq:ERR_R}
\end{equation}
where parameter $R$ is defined in \eqref{eq:Asymptototics_Lattice_L+-_zero}. The quantity $G_0a$ is equal to the energy released due to the breakage of one link between the oscillators whereas $G$ demonstrate the bulk change of energy. Apart from these quantities we also introduce the energy that would have been released if the horizontal links would break at potential breakage position $\eta=\eta_*$:
\begin{equation}
G_1=\frac{1}{2a}c_2(u_0(\eta_*+1)-u(\eta_*))^2.
\end{equation}

In the present paper we would like to analyse the effect of introduced anisotropy of elastic and strength properties. We define the parameter that characterise contrast in elastic properties:
\begin{equation}
\mu=\frac{c_2}{c_1}.
\end{equation}

\begin{figure}[h!]
\minipage{0.45\textwidth}
\center{\includegraphics[width=\linewidth] {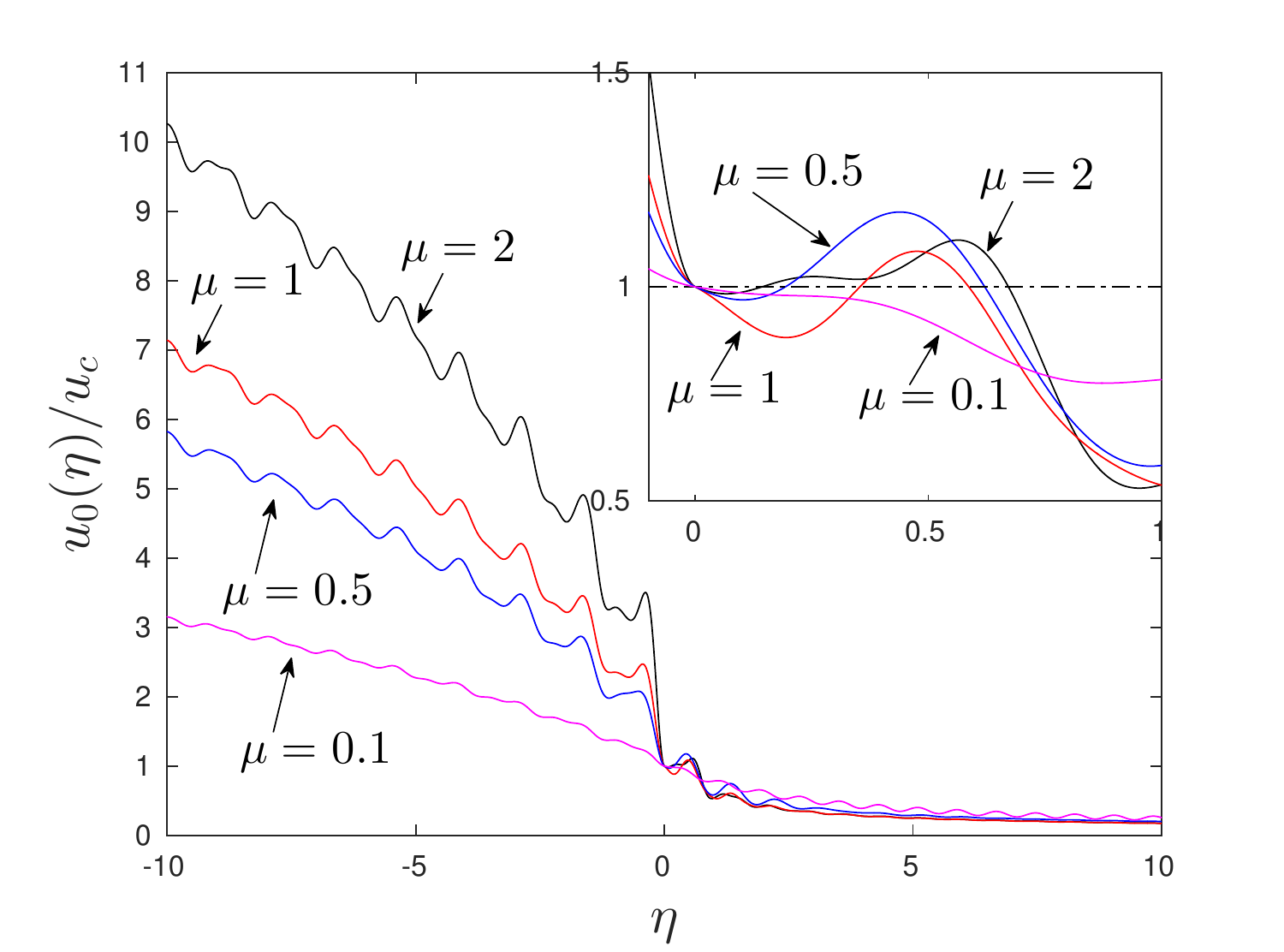} \\ a)}
\endminipage
\hfill
\minipage{0.45\textwidth}
\center{\includegraphics[width=\linewidth] {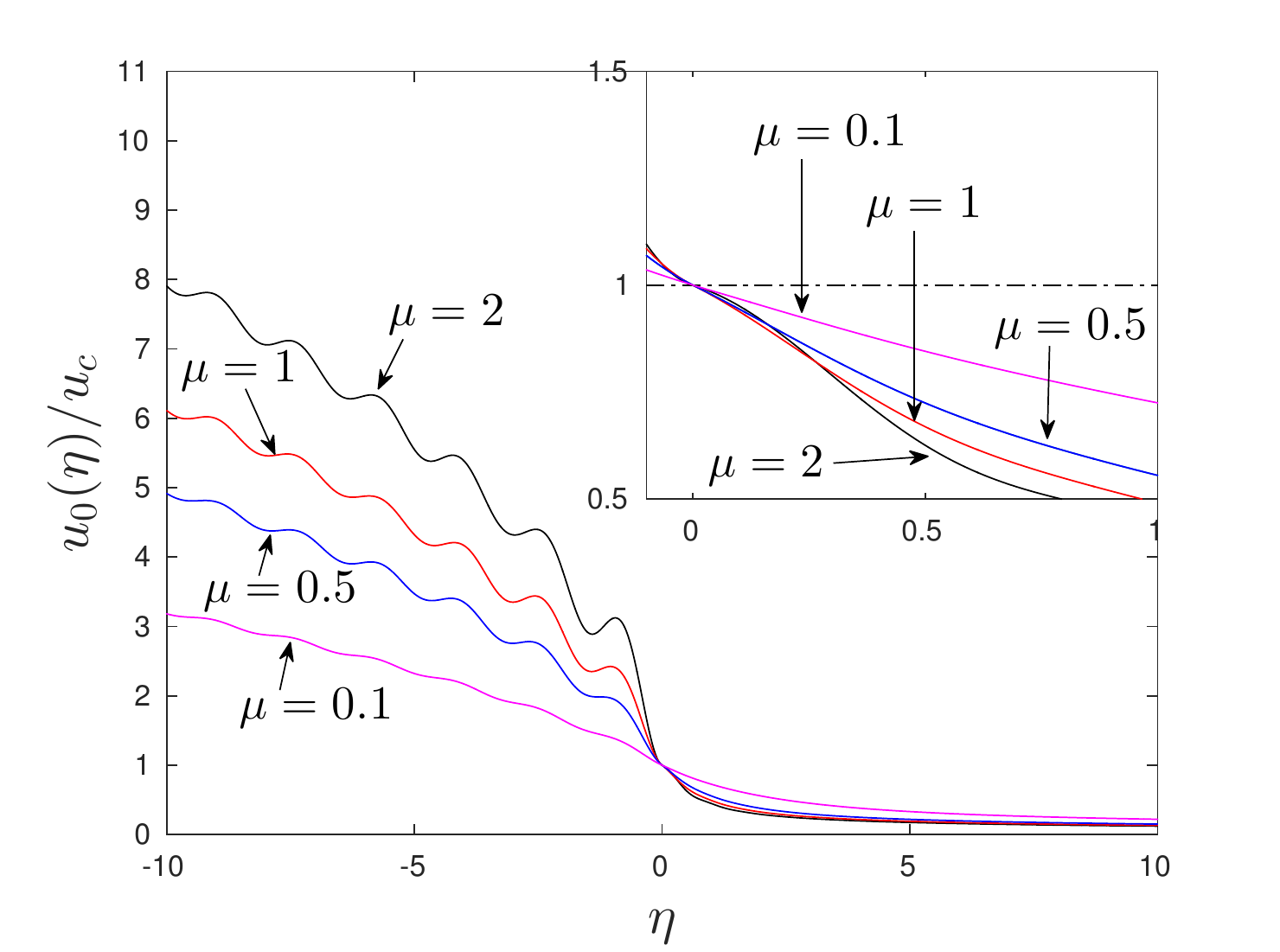} \\ b)}
\endminipage
\hfill
\captiondelim{. }
\caption[ ]{Displacement field $u_0(\eta)$ for different values of crack speed and $\mu$: a) $v=0.2v_c$, b) $v=0.5v_c$.}
\label{fig:Displacements}
\end{figure}

For the demonstration of the results we choose $M=1,c_1=1$ and values of $\mu=0.1,0.5,1,2$. The computations of the displacement fields require the prescribed values of $v$. The examples of displacement field $u_0(\eta)$ for different values of $\mu$ are shown in fig. \ref{fig:Displacements}.

After evaluation of displacement fields we can validate the second part of fracture criterion in \eqref{eq:FractureCondition_eta_1}. In the presented fig.\ref{fig:Displacements}a) we observe the fault of this criterion in the cases $\mu=0.5,1,2$ which makes these solutions unphysical for $v=0.2v_c$. At the same time the criterion \eqref{eq:FractureCondition_eta_1} is hold for every chosen value of $\mu$ at $v=0.5$ in fig. \ref{fig:Displacements}b). With these observations we define the following classes of the solutions:
\begin{itemize}
\item We call a regime of crack propagation at a certain speed {\it admissible} if conditions \eqref{eq:FractureCondition_eta} and \eqref{eq:FractureCondition_eta_1} are satisfied for the solution .
\item Otherwise, we call such regime {\it forbidden}.
\end{itemize}

Besides that we would also check the integrity of horizontal springs of stiffness $c_1$ within the admissible regimes. We perform such analysis by means of two different conditions. In the first condition:
\begin{figure}[h!]
\minipage{0.45\textwidth}
\center{\includegraphics[width=\linewidth] {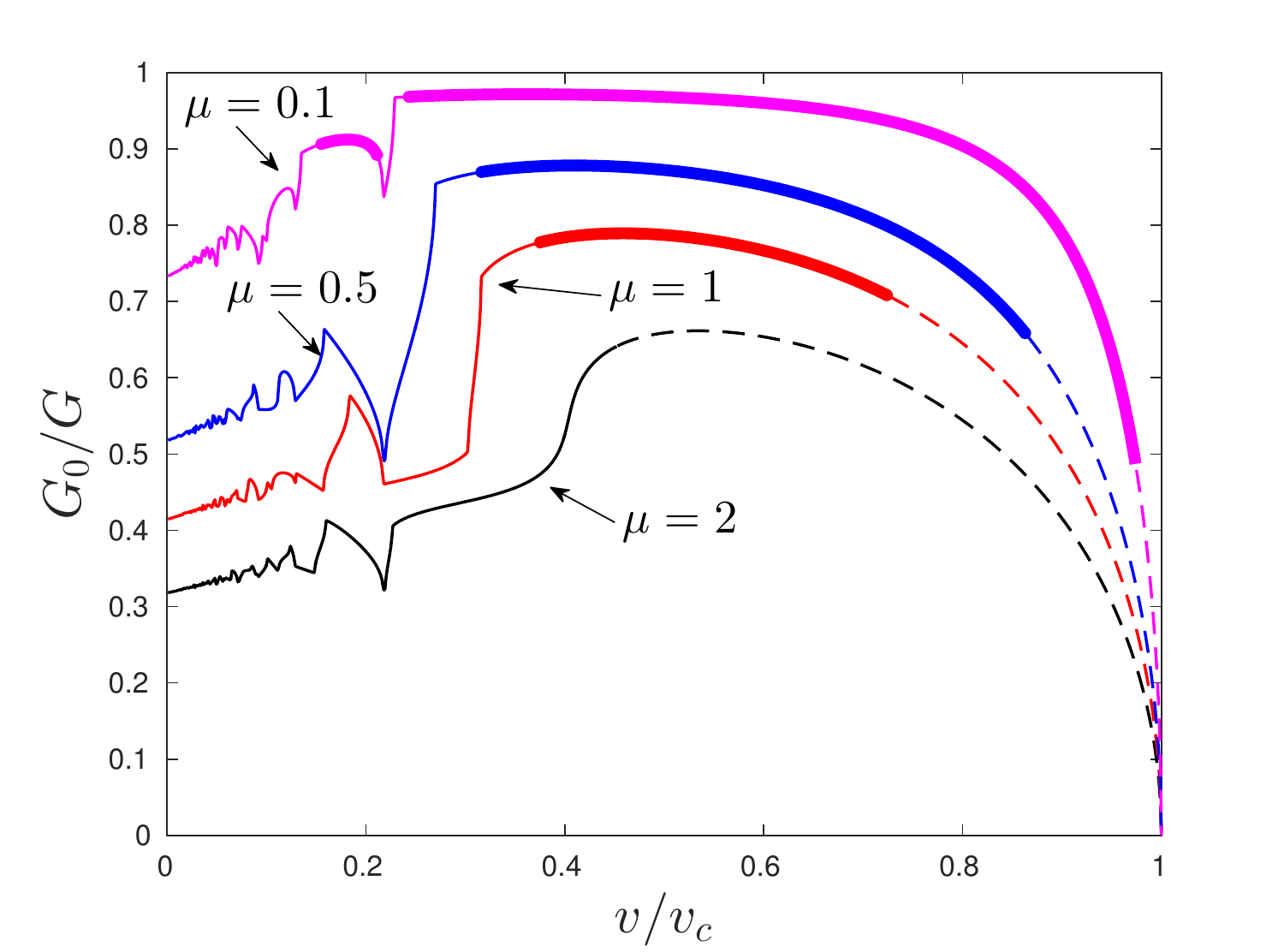} \\ a)}
\endminipage
\hfill
\minipage{0.45\textwidth}
\center{\includegraphics[width=\linewidth] {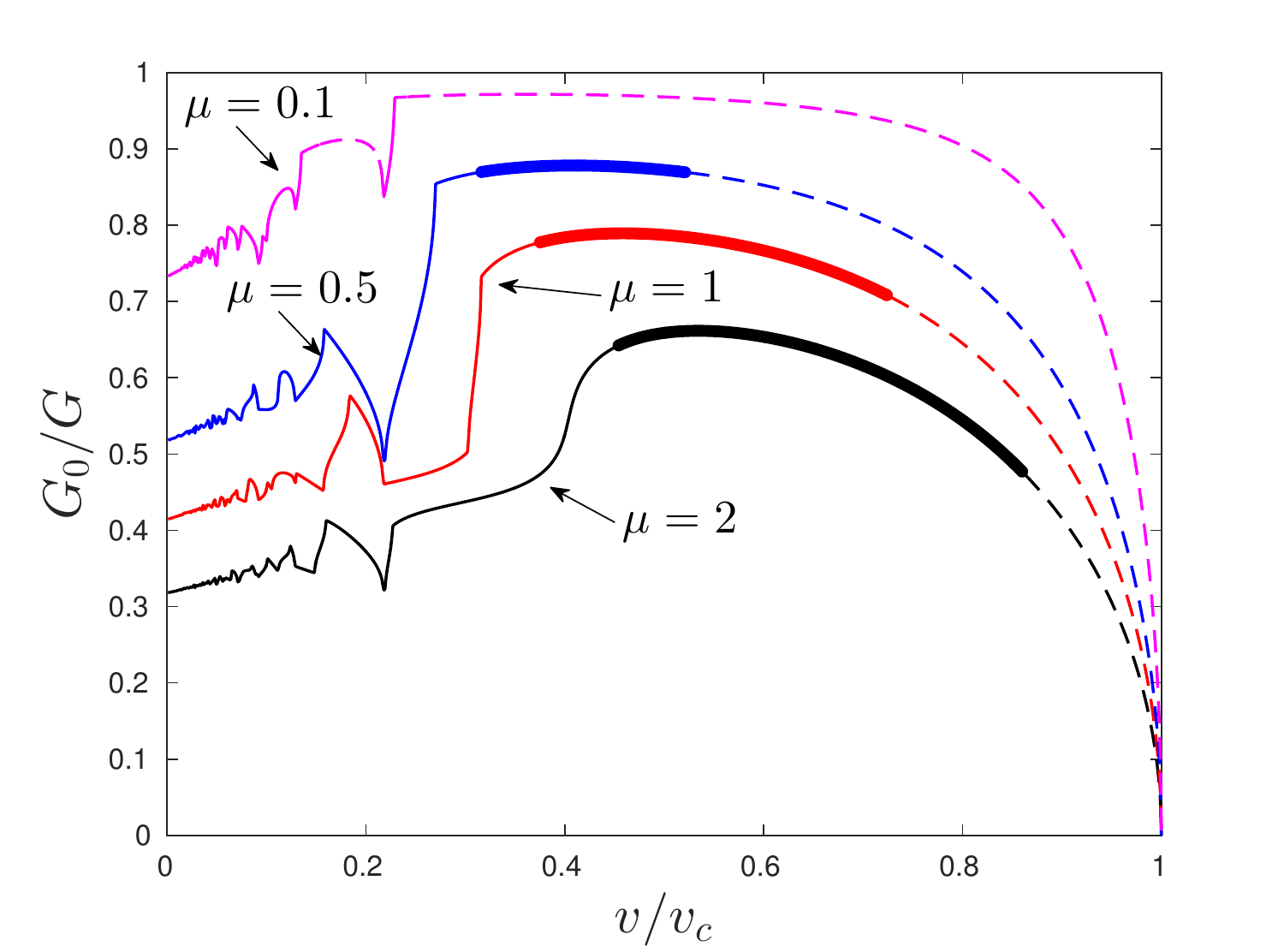} \\ b)}
\endminipage
\hfill
\captiondelim{. }
\caption[ ]{Dependence of energy release rate ratio $G_0/G$ on crack speed. Admissible regimes -- normal lines, forbidden regimes -- thick lines, dash-lines -- domains where the breakage of horizontal springs happen according to : a) condition \eqref{eq:Condition_1}, b) condition \eqref{eq:Condition_2}.}
\label{fig:ERR}
\end{figure}

\begin{equation}
\frac{G_0}{G_1}=1 \tag{C1}
\label{eq:Condition_1}
\end{equation}
we suppose the the strength in both directions is the same. For the second condition:

\begin{equation}
\frac{G_0}{G_1}=\mu \tag{C2}
\label{eq:Condition_2}
\end{equation}
we assume that the ratio released energy while fracture is proportional to the spring constant which is followed from the estimates of the theoretical strength. Posing either of these conditions allows to cut off the high values of the energy release rate as well as the high values of crack speed.

We perform the proceeding investigation of the obtained solution by means of definitions of admissible and forbidden regimes as well as the conditions \eqref{eq:Condition_1} and  \eqref{eq:Condition_1}. Similar analysis was carried out for a triangular cell lattice \cite{fineberg1999,pechenik2002,marder1995}, for a chain with anisotropic properties \cite{gorbushin2017} and non-local interactions \cite{gorbushin2016}. The complete analysis is shown in fig.\ref{fig:ERR} where the dependence of ratio $G_0/G$ on crack speed is presented according to \eqref{eq:ERR_R}. In this figure admissible regimes are marked with thick lines, forbidden regimes -- normal lines. Finally, the dash-line define the domains where condition \eqref{eq:Condition_1} (fig.\ref{fig:ERR}a)) or condition \eqref{eq:Condition_2} (fig.\ref{fig:ERR}b)) is failed.

As said in \cite{slepyan2012}, ratio $G_0/G$ demonstrate the fact that during the crack propagation not only the fracture energy (associated to term $G$) is released but also the elastic energy contained by the mechanical waves radiated by the crack tip. This fact is also observed in the dynamic fracture tests of materials \cite{rosakis1999}. That is to say, the lesser ratio $G_0/G$ is the more energy is carried by the fracture waves. Notice in fig. \ref{fig:ERR} that with increase of parameter $\mu$, characterising the lattice anisotropy, for a chosen value $v$, ratio $G_0/G$ decreases. This is reflected in the behaviour of $u_0(\eta)$ in fig. \ref{fig:Displacements} by the increase of amplitude of the waves with increase of $\mu$.

\ref{fig:Infinite Lattice}.
\begin{figure}[h!]
\center{\includegraphics[scale=0.5]{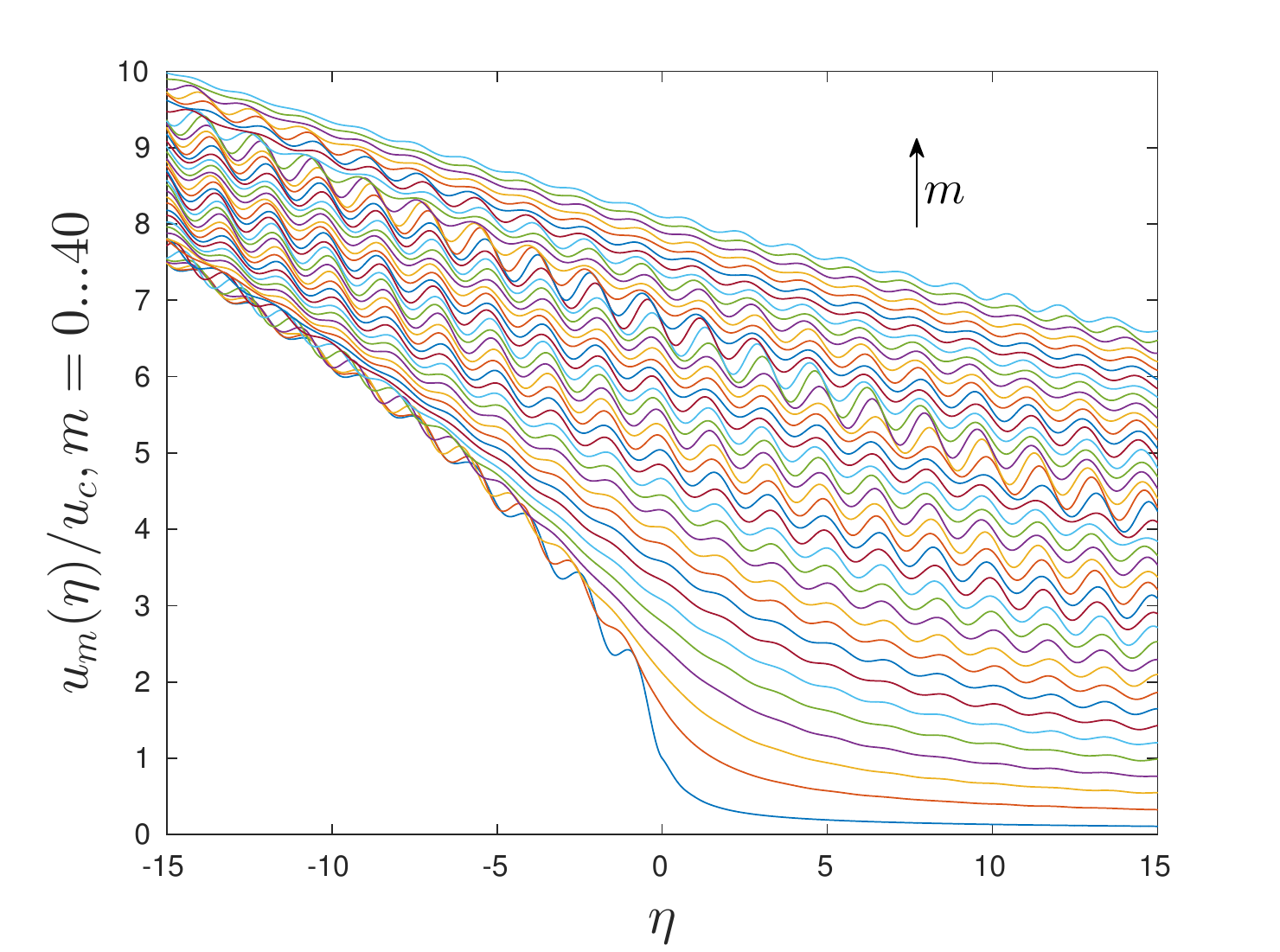}}
\captiondelim{. }
\caption[ ]{Displacements $u_m(\eta),m=0..40$ for $\mu=1,v=0.5v_c$.}
\label{fig:Displacements_layers}
\end{figure}

Non-monotonicity of $G_0/G$ leads to uncertainty in definition of an external load that causes fracture. Indeed, choosing a certain value of $G_0/G$ there can happen multiple intersections with the curves in fig. \ref{fig:ERR} which induce non-unique value of $v$. This ambiguousness may be avoided if the definite type of load is considered and its characteristic magnitude is plotted against a crack speed. Such dependences are shown for the cases of constant edge displacements \cite{fineberg1999,pechenik2002,marder1993}, constant strain or force at one ends of a chain \cite{slepyan2005} or a moving load \cite{gorbushin2017}.

The solution analysis reveal a qualitative effect of introduced material anisotropy. Particularly, with a significant contrast in the properties some separate intervals of admissible and forbidden regimes appear within a condition \eqref{eq:Condition_1}. This is vividly noticed for $\mu=0.1$ in fig. \ref{fig:ERR}a). This gives an opportunity of a crack propagation at low speeds which is not a case for isotropic lattice. In terms of \eqref{eq:Condition_1} the increase of parameter $\mu$ leads to the shrinkage of admissible domains and the possibility of failure of horizontal springs grows. This circumstance is related to the increase of released elastic energy with an increase of $\mu$. It finally bring a complete suppression of admissible regimes for $\mu=2$.

The consideration of condition \eqref{eq:Condition_2} effects only on the predictions of fracture of horizontal springs. In such a way for case $\mu=0.1$ the admissible domain is fully annihilated while such domain appears for $\mu=2$ (compare fig.\ref{fig:ERR}a) and fig.\ref{fig:ERR}b)). If condition \eqref{eq:Condition_1} takes place the domain of admissible regimes decreasing with an increase of $\mu$. This correlation is inverse if the condition \eqref{eq:Condition_1} is valid.

\begin{figure}[h!]
\minipage{0.45\textwidth}
\center{\includegraphics[width=\linewidth] {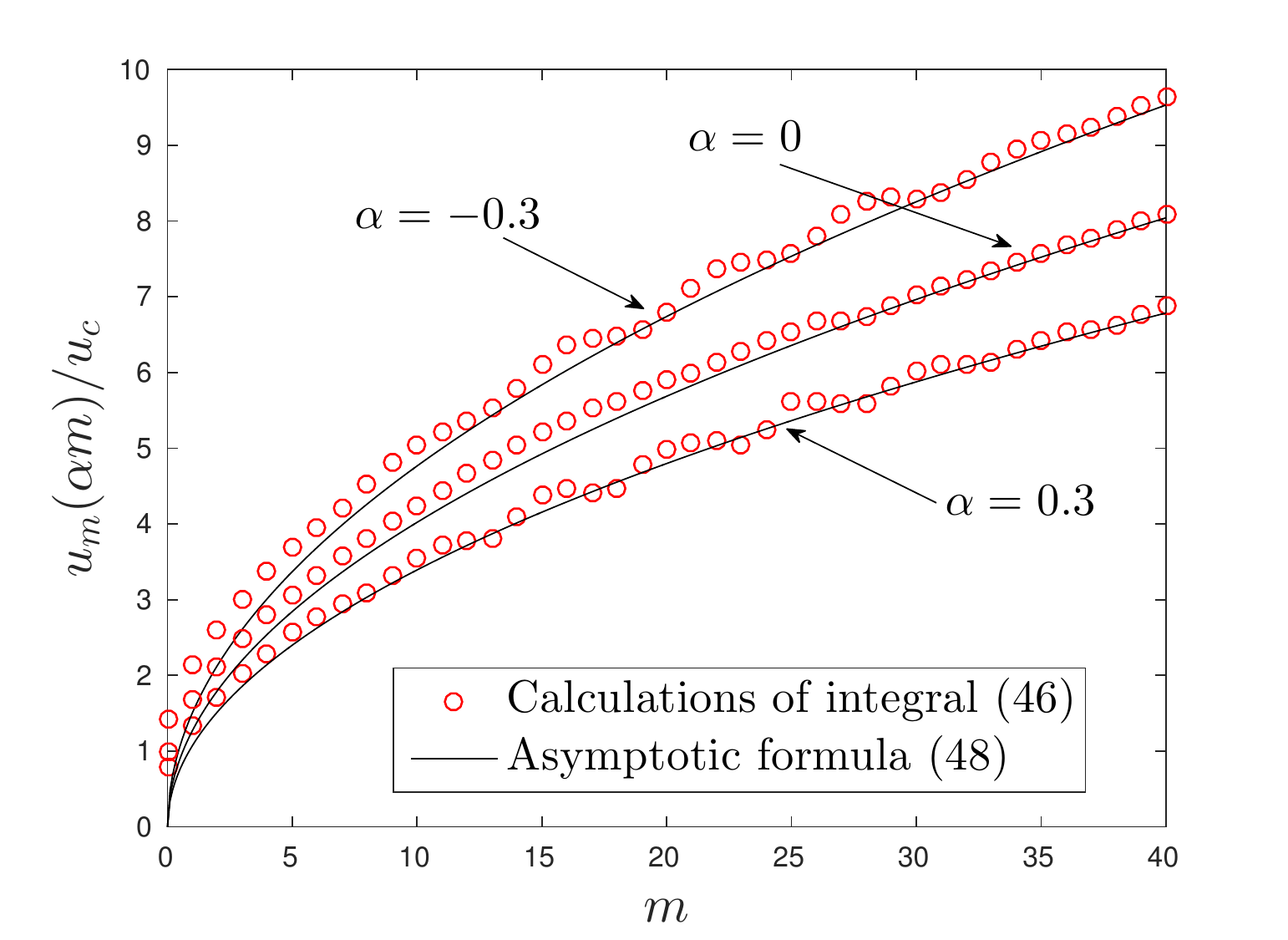} \\ a)}
\endminipage
\hfill
\minipage{0.45\textwidth}
\center{\includegraphics[width=\linewidth] {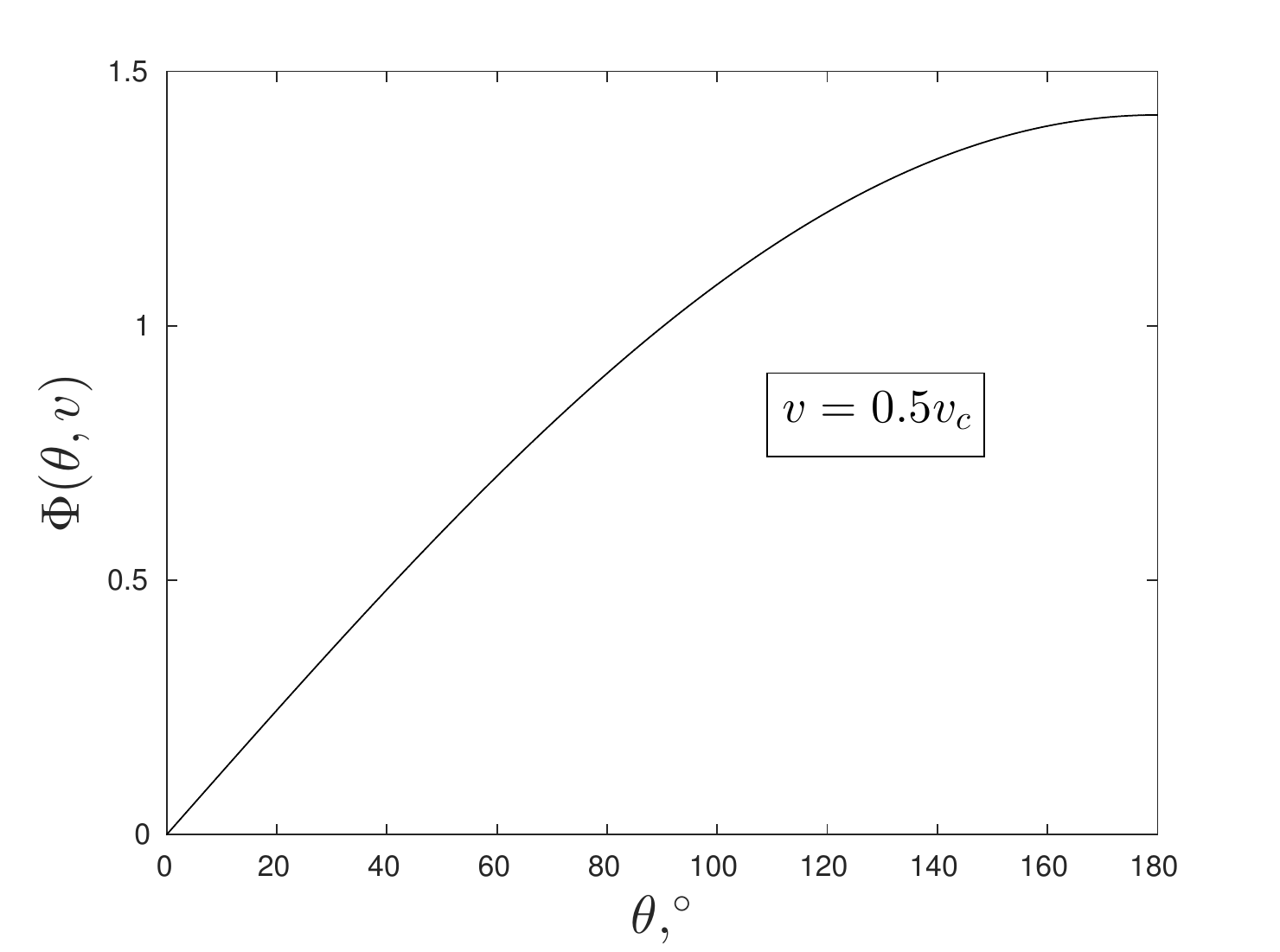} \\ b)}
\endminipage
\hfill
\captiondelim{. }
\caption[ ]{Examples in case of $v=0.5v_c$ and an isotropic lattice $v_c=\omega_0$: a) behaviour $u_m(\alpha m)$ according to \eqref{eq:Asymptotics_m}, where $\alpha=\cot(\theta)$, b) angle function $\Phi(\theta,v)$ in \eqref{eq:AngleFunction}.}
\label{fig:Asymptotics_m}
\end{figure}

We also present a plot of displacements $u_m(\eta),m=0..40$ for the case of isotropic lattice, $\mu=1$, and $v=0.5v_c$ in fig.\ref{fig:Displacements_layers}. Moreover, we display the figures for the behaviour of $u_m(\alpha m)$ when $m\to\infty$ in fig.\ref{fig:Asymptotics_m}a) and the angle function in fig.\ref{fig:Asymptotics_m}b) according to \eqref{eq:Asymptotics_m} and \eqref{eq:AngleFunction}, respectively. From these figures we conclude that the amplitude of the waves in each layer $m$ is different (see, fig.~\ref{fig:Displacements_layers}) and there is a tendency in slow growth of displacements in vertical direction which can be clearly seen \ref{fig:Asymptotics_m}a). This growth is greater in the region occupied by the crack ($\eta<0$) than ahead of it, which follows from the behaviour of angle function $\Phi(\theta,v)$ in \ref{fig:Asymptotics_m}b).

Additionally, we notice that the most dangerous elongations are found to be between layers $m=0$ and $m=1$ in the neighbourhood of a crack tip besides those layers adjacent to the symmetry line of a lattice. Nevertheless, these elongations do not exceed the threshold and remain to be less meaningful, in terms of fracture, than the elongations in horizontal direction. Additionally, we pay an attention to the fact that the displacement field in \ref{fig:Displacements_layers}, essentially, is a sum of two components: the waves radiated from a crack tip which amplitude is decaying with the distance from a crack tip and monotonic deformation which increases closer to infinity.

\section{Conclusions}

In the present work we considered the problem of straight crack propagation in an anisotropic square-cell lattice. One of the major result was the conclusion about the fact that the considerations of energetic characteristics of the problem are not enough for the reasonable predictions of stable crack propagation. The investigation of a solution is necessary.

The evaluation of displacement fields and their analysis allowed to distinguish two different regimes of crack propagation for a certain crack velocity -- admissible and forbidden regimes. Here it is worth of mentioning that constructed analytical solution is indeed reached which was demonstrated in \cite{gorbushin2017} for the one-dimensional chain of oscillators.

We also showed that the used approach can be effectively utilised for an analysis of integrity of horizontal springs in the lattice. For such study two different conditions were considered that were influenced by the contrast in elastic and strength properties in vertical and horizontal directions of a lattice. Particularly, we showed that for some values of material properties a steady-state crack propagation is not observed at all.

The introduced anisotropy allowed to discover a stable steady-state crack propagation at low velocities. This phenomenon is shown to be observed for high contrast in material properties. Moreover, the analysis allows to make predictions of crack velocities compatible with the stable crack movement.

Lastly, we revealed that any set of material parameters the failure of horizontal springs occurs at high values of crack speed. In such cases steady-state crack propagation can not be observed even within the admissible regimes.

\vspace{2mm} {\bf Acknowledgements} NG and GM acknowledge support from the EU projects CERMAT-2 (PITN-GA-2013-606878) and TAMER (IRSES-GA-2013-610547), respectively.
GM also acknowledges Royal Society of London for a partial support via Wolfson Research Merit Award.

\clearpage
\bibliographystyle{abbrv}
\bibliography{Bibliography}

\clearpage
\renewcommand{\theequation}{\Alph{section}.\arabic{equation}}
\numberwithin{equation}{section}
\appendix
\section{Appendix}
The obtaining of the asymptotic relations for lattice behaviour with the growth of $m$ takes into an account the behaviour of functions $U_0^{-}(k)$ and $\lambda(k)$ at zero. Indeed, when $m\to\infty$ the integration in the expressions for $u_m(\eta)$ is performed only within the intervals where $|\lambda(k)|=1$. However, the leading term of asymptotic behaviour of $u_m(\eta)$ when $m\to\infty$ is provided by the singularity encompassed in $U_0^{-}(k)$ when $k\to0$. To show this, let us subtract the leading terms of $U_0^{-}(k)$ and $\lambda(k)$ when $k\to\infty$ from the expressions for $u_m(\eta)$ and perform the analytical integration.

Firstly, we consider case $\eta>0$ and the integral to estimate is:
\begin{equation}
\int_{-\infty}^{\infty}(1-A\sqrt{0+ik}\sqrt{0-ik})^m\frac{a}{(0+ik)^{3/2}}e^{-ik\eta}\,d\eta,
\label{eq:Appendix_integral}
\end{equation}
where
\begin{equation*}
U^-(k)=\frac{a}{(0+ik)^{3/2}},\quad \lambda(k)=1-A\sqrt{0+ik}\sqrt{0-ik},\quad k\to0.
\end{equation*}

Constants $a,A>0$ are defined for the sake of reducing long expressions and will be substituted for the final expressions (see \eqref{eq:Asymptotics_Lambda_Zero} and \eqref{eq:Asymptotics_Lattice_Fourier}).  Integral \eqref{eq:Appendix_integral} can be computed using the technique of contour integration in the plane $k\in\mathbb{C}$. Notice, that according to the assumption $\eta>0$ the contour should be taken in half-plane $\Im k<0$. Moreover, in this plane function $\sqrt{0-ik}$ posses a branch cut along the imaginary axis. Thus, we construct contour $C=[-\rho,\rho]\cup\Gamma_1\cup[-i\rho,0]\cup[0,-i\rho]\cup\Gamma_2$ which consists of an interval along the real line, the arch of radius $\rho$ and $\Re k>0, \Im k<0$, two intervals along the branch cut of $\sqrt{0+ik}$ and the final arch of radius $\rho$ and $\Re k<0, \Im k<0$.

Notice that $\sqrt{0-ik}$ takes the following values along its cut:

\begin{equation*}
\sqrt{0-ik}=
\begin{cases}
-i\sqrt{z}, \quad \Re k\to0+,\\[2mm]
i\sqrt{z}, \quad \Re k\to0-,
\end{cases}
\end{equation*}
where $k=-iz,z>0$. At the same time $\sqrt{0+ik}=\sqrt{z}$. The Cauchy residue theorem allows to conclude that integration along contour $C$ the integrand in \eqref{eq:Appendix_integral} gives zero value. The integration along arches $\Gamma_1$ and $\Gamma_2$ also results in zero when $\rho\to\infty$. Thus, we get:
\begin{equation*}
\begin{gathered}
\int_{-\infty}^{\infty}(1-A\sqrt{0+ik}\sqrt{0-ik})^m\frac{a}{(0+ik)^{3/2}}e^{-ik\eta}\\
=-\frac{1}{i}\int_{\infty}^{0}(1+iAz)^m\frac{a}{z\sqrt{z}}e^{-z\eta}\,dz
-\frac{1}{i}\int_{0}^{\infty}(1-iAz)^m\frac{a}{z\sqrt{z}}e^{-z\eta}\,dz\\
=\int_{0}^{\infty}\frac{(1+iAz)^m-(1-iAz)^m}{iz\sqrt{z}}ae^{-z\eta}\,dz\\
=\int_{0}^{1/m}\frac{(1+iAz)^m-(1-iAz)^m}{iz\sqrt{z}}ae^{-z\eta}\,dz+
\int_{1/m}^{\infty}\frac{(1+iAz)^m-(1-iAz)^m}{iz\sqrt{z}}ae^{-z\eta}\,dz
\end{gathered}
\end{equation*}
Next, we are interested in the solutions along the ray:
$$\eta=\alpha m$$
Then, in the last equation of derivation we apply the change of variable, say $t=mz$. We then notice:
$$\left(1+iA\frac{t}{m}\right)^m\to e^{iAt},\quad \left(1-iA\frac{t}{m}\right)^m\to e^{-iAt},\quad m\to\infty$$
With this in mind we continue derivation:
\begin{equation*}
\begin{gathered}
\int_{0}^{1/m}\frac{(1+iAz)^m-(1-iAz)^m}{iz\sqrt{z}}ae^{-z\eta}\,dz+
\int_{1/m}^{\infty}\frac{(1+iAz)^m-(1-iAz)^m}{iz\sqrt{z}}ae^{-z\eta}\,dz\\
=a\sqrt{m}\int_{0}^{\infty}\frac{\left(1+iA\frac{t}{m}\right)^m-\left(1-iA\frac{t}{m}\right)^m}{t\sqrt{t}}e^{-\alpha t}\,dt\\
=2a\sqrt{m}\int_{0}^{\infty}\frac{\sin{At}}{t\sqrt{t}}e^{-\alpha t}\,dt
=2a\sqrt{m}\sqrt{A}\int_{0}^{\infty}\frac{\sin{x}}{x\sqrt{x}}e^{-(\alpha/A)x}dx\\
=2aA\sqrt{m}\frac{\sqrt{2\pi}}{\sqrt{\alpha+\sqrt{\alpha^2+A^2}}}
\end{gathered}
\end{equation*}

Finally, in case $\eta>0$, we achieve:
\begin{equation}
u_m(\alpha m)\sim\sqrt{\frac{2}{\pi}}\sqrt{\frac{v_c^2-v^2}{\omega_0^2}}\left(\frac{4\omega_0^2}{v_c^2-v^2}\right)^{1/4}\frac{u_c}{R}\frac{\sqrt{m}}{\sqrt{\alpha+\sqrt{\alpha^2+\frac{v_c^2-v^2}{\omega_0^2}}}},\quad \alpha>0,\quad m\to\infty,
\end{equation}
where used notations introduced in the main part of the article. Let us now turn to the situation when $\eta<0$. The integral in \eqref{eq:Appendix_integral} can be written as:
\begin{equation}
\begin{gathered}
\int_{-\infty}^{\infty}(1-A\sqrt{0+ik}\sqrt{0-ik})^m\frac{a}{(0+ik)^{3/2}}e^{-ik\eta}\,d\eta\\
=\int_{-\infty}^{\infty}\left[(1-A\sqrt{0+ik}\sqrt{0-ik})^m-1\right]\frac{a}{(0+ik)^{3/2}}e^{-ik\eta}\,d\eta+\int_{-\infty}^{\infty}\frac{a}{(0+ik)^{3/2}}e^{-ik\eta}\,d\eta
\end{gathered}
\label{eq:Appendix_integral_2}
\end{equation}
The evaluation of the second integral gives a straightforward result by means of contour integration:
\begin{equation}
\int_{-\infty}^{\infty}\frac{a}{(0+ik)^{3/2}}e^{-ik\eta}\,d\eta=4a\sqrt{-\pi\eta},\quad \eta<0
\label{eq:Appendix_integral_eta_neg}
\end{equation}

Notice, that in this case we need consider contour in half-plane $\Im k>0$ where $\sqrt{0+ik}$ possesses a cut:
\begin{equation*}
\sqrt{0+ik}=
\begin{cases}
i\sqrt{z}, \Re k\to0+,\\
-i\sqrt{z}, \Re k\to0-,
\end{cases}
\end{equation*}
where $k=iz,\quad z>0$, whereas $\sqrt{0-ik}=\sqrt{z}$. The reasoning concerning the construction of contour integration is similar to the previous case. The application of the Cauchy residue theorem and Jordan's lemma leads to:
\begin{equation*}
\begin{gathered}
\int_{-\infty}^{\infty}\left[(1-A\sqrt{0+ik}\sqrt{0-ik})^m-1\right]\frac{a}{(0+ik)^{3/2}}e^{-ik\eta}\\
=\int_{\infty}^{0}\left[(1+iAz)^m-1\right]\frac{a}{z\sqrt{z}}e^{z\eta}\,dz
-\int_{0}^{\infty}\left[(1-iAz)^m-1\right]\frac{a}{z\sqrt{z}}e^{z\eta}\,dz\\
=\int_{0}^{\infty}\frac{2-(1+iAz)^m-(1-iAz)^m}{z\sqrt{z}}ae^{z\eta}\,dz\\
=\int_{0}^{1/m}\frac{2-(1+iAz)^m-(1-iAz)^m}{iz\sqrt{z}}ae^{z\eta}\,dz+
\int_{1/m}^{\infty}\frac{2-(1+iAz)^m-(1-iAz)^m}{iz\sqrt{z}}ae^{z\eta}\,dz
\end{gathered}
\end{equation*}

Again, we make a change of variable $t=mz$ and assume that:
$$\eta=-\alpha m.$$
Taking the limit $m\to\infty$ we arrive to the expression:
\begin{equation}
\begin{gathered}
\int_{0}^{1/m}\frac{2-(1+iAz)^m-(1-iAz)^m}{iz\sqrt{z}}ae^{z\eta}\,dz+
\int_{1/m}^{\infty}\frac{2-(1+iAz)^m-(1-iAz)^m}{iz\sqrt{z}}ae^{z\eta}\,dz\\
=a\sqrt{m}\int_{0}^{\infty}\frac{1-\cos{At}}{t\sqrt{t}}e^{-\alpha t}\,dt
=2a\sqrt{m}\sqrt{A}\int_{0}^{\infty}\frac{1-\cos{x}}{x\sqrt{x}}e^{-(\alpha/A)x}\,dx\\
=4a\sqrt{\pi m}\left(-\sqrt{\alpha}+(\alpha^2+A^2)^{1/4}
\cos{\left(\frac{\arccot{\frac{\alpha}{A}}}{2}\right)}\right),
\end{gathered}
\end{equation}
where $\arccot{x}$ is the inverse cotangent function. Sum of the last result with \eqref{eq:Appendix_integral_eta_neg} provides the final for of asymptotic behaviour:

\begin{equation*}
u_m(-\alpha\eta)\sim\frac{2u_c}{\sqrt{\pi}}\frac{1}{R}\left(\frac{4\omega_0^2}{v_c^2-v^2}\right)^{1/4}\left(\alpha^2+\frac{v_c^2-v^2}{\omega_0^2}\right)^{1/4}\cos{\left(\frac{\arccot{\left(-\alpha\sqrt{\frac{\omega_0^2}{v_c^2-v^2}}\right)}}{2}\right)}\sqrt{m},\quad m\to \infty, \alpha>0
\end{equation*}
One can simplify the trigonometric term in the last expression using the following expressions:
\begin{equation*}
\arccot{\left(\frac{1}{x}\right)}=\arctan{(x)},\quad \arctan{(x)}=2\arctan\left(\frac{x}{1+\sqrt{x^2+1}}\right),\quad \cos{(\arctan(x))}=\frac{1}{\sqrt{1+x^2}}
\end{equation*}
to obtain:
\begin{equation*}
u_m(-\alpha\eta)\sim\sqrt{\frac{2}{\pi}}\sqrt{\frac{v_c^2-v^2}{\omega_0^2}}\left(\frac{4\omega_0^2}{v_c^2-v^2}\right)^{1/4}\frac{u_c}{R}
\frac{\sqrt{m}}{\sqrt{-\alpha+\sqrt{\alpha^2+\frac{v_c^2-v^2}{\omega_0^2}}}},\quad \alpha>0,\quad m\to\infty
\end{equation*}
Notice, that the last expression is of the same form as in the previously considered case $\eta<0$. Thus, we can present final formula of asymptotic expression in a general form:
\begin{equation}
u_m(\alpha m)\sim \frac{2u_c}{R}\left(\frac{4\omega_0^2}{v_c^2-v^2}\right)^{1/4}\sqrt{\frac{m}{2\pi}}\left(\dfrac{v_c^2-v^2}{\omega_0^2}\frac{1}{\alpha+\sqrt{\alpha^2+\frac{v_c^2-v^2}{\omega_0^2}}}\right)^{1/2},\quad m\to\infty
\label{eq:App_Asymptotics_m}
\end{equation}

One can observe that introducing the radius vector $\rho=\sqrt{m^2+\eta^2}=m\sqrt{1+\alpha^2}$ and the angle $\theta$, such that $\cot\theta=\alpha$ (and, thus, $m=\rho\sin\theta$,
$\eta=\rho\cos\theta$), the representation \eqref{eq:Asymptotics_m}
can be conveniently rewritten in the following manner:
\begin{equation}
u_m(\eta)\sim \frac{2u_c}{R}\left(\frac{4\omega_0^2}{v_c^2-v^2}\right)^{1/4}\!\!\sqrt{\frac{\rho}{2\pi}}\,\,\,\Phi(\theta,v),\quad \rho\to\infty,
\label{eq:Asymptotics_rho}
\end{equation}
\begin{equation}
\Phi(\theta,v)=
\left(\sqrt{\cos^2\theta+\frac{v_c^2-v^2}{\omega_0^2}\sin^2\theta}- \cos\theta \right)^{1/2},\quad 0\le \theta<\pi,
\label{eq:AngleFunction_2}
\end{equation}
In the case of a static problem ($v=0$) and an isotropic lattice ($v_c=\omega_0$) we have:
\[
\Phi(\theta,0)=\sin\frac{\theta}{2},\quad 0\le \theta<\pi,
\label{eq:AngleFunction_3}
\]
that correspond to the asymptotic solution near the crack tip in homogeneous material under the Mode III deformation.

Using the relationships \eqref{eq:ERR_R}, asymptotics \eqref{eq:Asymptotics_rho} can be rewritten in an equivalent form:
\begin{equation}
u_m(\eta)\sim 2\sqrt{\frac{aG}{c_1}}\left(\frac{\omega_0^2}{v_c^2-v^2}\right)^{1/4}\!\!\sqrt{\frac{\rho}{2\pi}}\,\,\,\Phi(\theta,v),\quad \rho\to\infty.
\label{eq:Asymptotics_rho_G}
\end{equation}

It is worth of mentioning that continuum formulation similar problem with moving crack $v>0$ under mode III in the case of homogeneous material leads to the following asymptotic behaviour of the displacement field at the crack tip (see e.g. \cite[p.356-360]{broberg1999}):
\begin{equation*}
u(r,\theta)\sim 2\sqrt{\frac{\mathcal{G}}{\mu}}\left(1-\frac{v^2}{v_c^2}\right)^{-1/4}\sqrt{\frac{r}{2\pi}}\left(\sqrt{1-\frac{v^2}{v_c^2}\sin^2{\theta}}-\cos{\theta}\right),\quad r\to0,
\end{equation*}
where $\mu$ -- shear modulus, $\mathcal{G}$ -- macroscopic energy release rate and $v_c$ is a shear wave speed in this case. Thus, we the considered microscopic solution reflects the behaviour of the macroscopic case.
\end{document}